\documentclass[letterpaper, 10 pt, conference]{ieeeconf}  

\IEEEoverridecommandlockouts                              
\usepackage{amsfonts}
\usepackage{graphicx}
\usepackage[dvipsnames]{xcolor}
\usepackage{amsmath}
\usepackage{amssymb}
\usepackage{dsfont}

\usepackage{ifthen}
\usepackage{color}
\usepackage{cite}

\usepackage{mathtools}
\usepackage{booktabs}
\usepackage{url}
\usepackage{hyperref}
\usepackage[font={small}]{caption}
\usepackage{subfig}
\usepackage{float}
\usepackage{multirow}
\usepackage{comment}
\usepackage{tikz}
\usepackage{pgfplots}
\pgfplotsset{compat=1.12}
\usepackage[ruled,vlined]{algorithm2e}

\usepackage{braket}
\usepackage{physics}

\usepackage{stmaryrd}

\usepackage{cuted}
\usepackage{bm}

\def\scr#1{{\cal #1}}
\newcommand{\R}{{\rm I\!R}}

\newcommand{\bbb}{\mathbb}

\newtheorem{lemma}{Lemma}

\newtheorem{assumption}{Assumption}

\newcommand{\1}{\mathbf{1}}

\DeclareMathOperator*{\argmin}{arg\,min}

\newcommand{\diag}{{\rm diag}}

\newcommand{\dbb}[1]{\llbracket #1 \rrbracket}

\newcommand{\Ker}{\text{ker}}

\title{\LARGE \bf Distributed Variational Quantum Linear Solver
}

\author{Tong Shen \hspace{.3in} Zeru Zhu \hspace{.3in} Ji Liu
\thanks{T. Shen and Z. Zhu are with the Department of Applied Mathematics and Statistics at Stony Brook University.
J.~Liu is with the Department of Electrical and Computer Engineering at Stony Brook University.
Email addresses: \href{mailto:tong.shen.1@stonybrook.edu}{tong.shen.1@stonybrook.edu}, 
\href{mailto:zeru.zhu@stonybrook.edu}{zeru.zhu@stonybrook.edu},\;\href{mailto:ji.liu@stonybrook.edu}{ji.liu@stonybrook.edu}.
}
}

\begin{document}

\maketitle
\thispagestyle{empty}
\pagestyle{empty}

\begin{abstract}
This paper develops a distributed variational quantum algorithm for solving large-scale linear equations. For a linear system of the form $Ax=b$, the large square matrix $A$ is partitioned into smaller square block submatrices, each of which is known only to a single noisy intermediate-scale quantum (NISQ) computer. Each NISQ computer communicates with certain other quantum computers in the same row and column of the block partition, where the communication patterns are described by the row- and column-neighbor graphs, both of which are connected. The proposed algorithm integrates a variant of the variational quantum linear solver at each computer with distributed classical optimization techniques. The derivation of the quantum cost function provides insight into the design of the distributed algorithm. Numerical quantum simulations demonstrate that the proposed distributed quantum algorithm can solve linear systems whose size scales with the number of computers 
and is therefore not limited by the capacity of a single quantum computer.
\end{abstract}

\section{Introduction}

Quantum computing algorithms have shown great promise of substantial speedup over their classical counterparts; notable examples include solving linear systems of equations \cite{3_harrow2009hhl}, integer factorization \cite{01_shor1994factorization}, search of unstructured databases \cite{grover1996fast}, and successful experimental demonstrations by Google \cite{02_arute2019quantum}. However, current noisy intermediate-scale quantum (NISQ) computers are prone to errors due to the limited number of qubits, restricted qubit
connectivity, and short qubit coherence times, which necessitates error correction in quantum computation \cite{10_preskill2018nisq}.
 Unfortunately, realizing a fault-tolerant quantum computer requires many technological breakthroughs; IBM estimates that such systems may become feasible within the next decade \cite{03_IBMQuantumRoadmap2023}. With the advent of quantum computing, there has also been increasing interest in the quantum internet, in which quantum computers are connected through quantum channels that exploit entanglement to enable communication over long distances with enhanced security. Although recent experiments have demonstrated multi-node quantum networks \cite{04_liu2024creation} and entanglement distribution between quantum memories over 50 kilometers of optical fiber \cite{05_yu2020entanglement}, reliable long-distance quantum communication capable of transmitting quantum states between remote quantum computers is still a distant dream.

With the above in mind, a natural question is how to utilize the current NISQ era quantum computers and classical communication to explore the maximum potential of quantum computing and solve large-scale realistic problems in the coming decade, until fault-tolerant quantum computers become available. The answer is promising because of the following two key factors. 

First, recent literature suggests that a class of algorithms, known as variational quantum algorithms (VQAs), may offer potential quantum advantage in the NISQ era \cite{mcclean2016theory}. These algorithms mitigate the limitations imposed by short qubit coherence times by employing a hybrid quantum-classical framework, in which shallow quantum circuits are executed on quantum processors while parameter optimization is performed on classical computers. Such hybrid structures can be naturally integrated with classical communication networks. 
Second, over the past two decades, the efforts of the control and optimization communities have led to the birth of the fields of distributed (classical) control and optimization, which have achieved great success in both theory and practice and have found a wide range of applications including robotic teams, sensor networks, and electric power grids \cite{molzahn2017survey}. Compared with traditional single-agent systems, distributed multi-agent systems have the advantages of fault tolerance, low cost, scalability, and privacy preservation, and can thus solve problems that are computationally expensive or even impossible for a single agent to solve. 

Solving systems of linear equations is probably the most fundamental computational task in science and engineering. Motivated by the limitations of single NISQ computers and conventional VQAs, together with the advantages of classical distributed multi-agent algorithms, this paper develops the first distributed VQA for solving large-scale linear equations. 
The proposed algorithm integrates a variant of the variational quantum linear solver executed on each NISQ computer with distributed classical optimization techniques coordinated through classical communication.
Numerical quantum simulations demonstrate that the proposed distributed
quantum algorithm can solve linear systems whose size scales with the
number of quantum computers and is therefore no longer limited by the capacity of a single quantum computer. These results highlight the potential of distributed VQAs for large-scale quantum computing.

We begin with the quantum linear system problem to help readers become familiar with the relevant quantum computing background.

\subsection{Quantum Linear System Problem}

An $n$-qubit quantum state is represented by a normalized vector in a $2^n$-dimensional Hilbert space.
The quantum linear system problem aims to prepare a quantum state proportional to the solution vector of a linear system $Ax=b$,
where $A$ is a given real matrix and $b$ is a given real vector.
Without loss of generality, it is assumed that $A$ is a square matrix of size $2^n\times 2^n$ and that $\|b\|=1$, where $\|\!\cdot\!\|$ denotes the Euclidean norm induced by the Hilbert space inner product. 
If $A$ is not of size $2^n\times2^n$, the equation can be embedded into an equivalent larger square linear system by augmenting the matrix and vector with zeros.
The dimension $2^n$ corresponds to the state space of an $n$-qubit quantum system, and the
normalization condition ensures that $b$ can be encoded as a quantum state
$|b\rangle=\sum_{i=0}^{2^n-1} b_i|i\rangle$, where $|i\rangle$ denotes the $i$th computational basis state of the $n$-qubit system.

The seminal Harrow-Hassidim-Lloyd (HHL) algorithm was the first quantum algorithm for solving linear systems and established that a quantum computer can prepare a quantum state proportional to the solution vector under suitable assumptions, but it relies on phase estimation and related subroutines that lead to substantial circuit depth \cite{3_harrow2009hhl}.
Subsequent algorithms improved the dependence of the runtime on the condition number and precision compared with HHL, leading to nearly optimal quantum linear system solvers using more advanced techniques such as block encoding \cite{5_childs2017precision}, quantum singular value transformation \cite{6_gilyen2019qsvt}, and adiabatic-inspired approaches \cite{43_Yigit2019adiabatic}. 
However, these methods are primarily designed for the fault-tolerant quantum computing regime, where deep circuits can be executed reliably. 
In contrast, the variational quantum linear solver (VQLS) is designed for near-term hybrid 
quantum-classical implementation, using relatively shallow parameterized circuits and 
thus offering better compatibility with current NISQ devices \cite{vqls}. 
For this reason, we focus on VQLS in this paper. 
It is worth emphasizing that the runtime of VQLS depends on the classical optimization process and ansatz design, and therefore it does not provide the same worst-case complexity guarantees as those fault-tolerant quantum linear system algorithms.

\subsection{Variational Quantum Linear Solver}

VQLS is a hybrid quantum-classical algorithm for solving the quantum linear system problem, in which a quantum processor prepares parameterized quantum states and estimates the cost function via measurements, while a classical computer updates the parameters using an optimization routine.
Specifically, VQLS prepares a parameterized state 
$\ket{x(\theta)} = V(\theta)\ket{0}$ using a variational ansatz, where $\theta$ is a vector of variational parameters defining the parameterized quantum circuit $V(\theta)$.
The goal is to approximate a normalized quantum state proportional to the solution of the linear system $Ax=b$ by iteratively updating the parameter vector $\theta$. 
Starting from an initial parameter vector $\theta (0)$, the quantum processor prepares the state $\ket{x(\theta)}$ and evaluates a carefully designed cost function $C(\theta)$ through quantum measurements. 
A cost function is constructed to measure how well the state 
$A|x(\theta)\rangle$ aligns with $|b\rangle$.
The measured value of $C(\theta)$ is then sent to a classical optimizer, 
which updates the parameter vector according to
$
\theta (k+1) = \theta (k) + \Delta\theta(k)$,
where $\Delta\theta(k)$ is determined by the chosen optimization 
routine (e.g., gradient-based or gradient-free methods). This process 
is repeated until the cost function falls below a desired threshold, 
indicating that the prepared state $\ket{x(\theta)}$ approximates the 
normalized solution of the linear system.
Further details of the original VQLS algorithm can be found in \cite{vqls}. 


The original VQLS formulation in \cite{vqls} assumes that the linear system $Ax=b$ has a unique solution $A^{-1}b$. While the variational framework can in principle
be applied to systems with multiple solutions or to least squares problems through suitable reformulations, these cases are not covered by the original analysis.
Moreover, when $Ax=b$ has a unique solution $x^\star$, $\|Ax^\star\|=\|b\|$. After VQLS prepares a normalized state $\ket{x}$ proportional to $x^\star$, the solution vector can be easily recovered by multiplying $\ket{x}$ by a normalization factor determined from $\|b\|$. In contrast, for a 
least squares solution $x^*$, 
$Ax^*\neq b$ and $\|Ax^*\|$ is unknown. Consequently, recovering $x^*$ from a normalized quantum state proportional to it requires estimating the corresponding normalization factor, which is nontrivial.
Note that the matrix $A$ in VQLS is 
assumed to admit an efficient linear combination of unitaries
(LCU) representation
$A=\sum_{k} c_k A_k$, 
where $c_k$ are coefficients and $A_k$ are unitary matrices. This assumption allows $A$ to be implemented on a quantum computer using standard LCU techniques. It is worth mentioning that obtaining such a decomposition may be computationally challenging in practice, especially when $A$ is large-scale.

The VQLS paper \cite{vqls} evaluates the cost function using Hadamard test based circuits at the algorithmic level. 
However, its large-scale numerical experiments simulate linear systems of dimension up to $2^{50}\times 2^{50}$ using a matrix product state simulator, which directly simulates the quantum circuit evolution rather than explicitly implementing Hadamard test circuits. Explicit Hadamard test evaluation requires computing many expectation 
values term by term and therefore becomes computationally prohibitive for such large-scale classical simulations.
This computational limitation naturally suggests a multi-agent extension in which each agent evaluates a smaller number of
expectation terms in parallel. 


\subsection{Multi-Agent Quantum Computing}

Multi-agent quantum computing has been explored for several quantum machine learning tasks, including federated and consensus-based distributed training settings, where agents coordinate through classical or quantum communication \cite{24_chen2021fqml,25_chehimi2021qfl,du2022tqe,26_chen2024cdqkl,27_swaminathan2024securekernel,22_barral2025review}. However, multi-agent quantum algorithms for solving linear systems have received little attention. Indeed, to the best of our knowledge, such algorithms have not yet been developed, although some works suggest that multi-agent implementations for the quantum linear system problem may be possible without providing detailed algorithmic descriptions.
Motivated by the preceding discussion, this paper proposes a distributed multi-agent version of VQLS. 

Related research directions include quantum circuit partitioning \cite{circuit_partitioning} and circuit cutting \cite{29_peng2020simulating}, which aim to decompose a large quantum circuit into smaller subcircuits so that large-scale quantum computations can be executed on limited hardware resources. These approaches are fundamentally different from the distributed coordination framework considered in this paper.
Circuit partitioning and cutting methods first design a large circuit for a target computational task and then divide it into smaller pieces for implementation, where the decomposition is largely determined by the circuit structure and therefore not freely controllable. In contrast, the distributed coordination approach partitions the computational task itself and assigns smaller information subsets to different agents, which subsequently coordinate by exchanging processed information.

\section{Distributed Multi-Agent VQLS}

We first describe how the information of a large-scale linear system is distributed across a multi-agent network. 

\subsection{Problem Formulation}

Let each agent be a $q$-qubit quantum computer capable of representing vectors of dimension $2^q$, and accordingly partition the $2^n \times 2^n$ real matrix $A$ and the $2^n$-dimensional real vector $b$ as
\begin{align*}
    A = \left[ 
    \begin{array}{ccc}
        A_{11} & \cdots & A_{1m}  \\
        \vdots & \ddots & \vdots  \\
        A_{m1} & \cdots & A_{mm}  
    \end{array}
    \right] \;\;\; \text{and}
 \;\;\;
    b = \left[ 
    \begin{array}{c}
        b_{1} \\
        \vdots \\
        b_{m}
    \end{array}
    \right],
\end{align*}
where $m=2^{n-q}$, each $A_{ij}$ is a submatrix of size $2^q \times 2^q$, and each subvector $b_i$ has dimension $2^q$.
Each agent is assigned exactly one submatrix $A_{ij}$, and for this reason, the index $\dbb{ij}$ is used to identify the corresponding agent. The total number of agents is therefore $m^2$. 
Each agent $\dbb{ij}$ is also assigned a vector $b_{ij}$ of dimension $2^q$ such that $\sum_{j=1}^m b_{ij} = b_i$. 
The goal of each agent $\dbb{ij}$ is to compute a vector $x_j$ of dimension $2^q$ such that their stacked vector
$$    x^* = \left[ 
    \begin{array}{c}
        x_1^* \\
        \vdots \\
        x_m^*
    \end{array}
    \right] \in\; \argmin_{x\in\R^{2^n}} \|Ax-b\|^2. $$ 
That is, $x^*$ is a least squares solution to $Ax=b$.
In this way, agents $\dbb{ij}$, $i\in\{1,\ldots,m\}$ with a fixed $j$ reach a consensus on the corresponding subvector $x_j^*$ of a least square solution.
Meanwhile, agents $\dbb{ij}$, $j\in\{1,\ldots,m\}$ with a fixed $i$ need to coordinate, as their goals are coupled.

Distributed classical linear system problems have been studied in the systems and control community for over a decade \cite{tacle}, including the block partition setting described above, as well as more general partition structures \cite{wang2020scalable,huang2022scalable,huang2024distributed,pham2023distributed}. 
Most existing algorithms for the block partition setting are based on continuous-time differential equations and are therefore not directly compatible with quantum computing due to the fundamentally discrete and unitary nature of quantum computation.
Although some works \cite{huang2024distributed,pham2023distributed} consider discrete-time iterative algorithms,
these methods still do not readily extend to the hybrid quantum-classical framework because of the fundamentally different information representation and processing mechanisms.

To present our algorithm and formalize the distributed setting, we introduce the following communication structures among the agents. For each agent $\dbb{ij}$, agent $\dbb{ik}$ is called a row-neighbor of agent $\dbb{ij}$ if it can exchange information with agent $\dbb{ij}$. Similarly, agent $\dbb{kj}$ is called a column-neighbor of agent $\dbb{ij}$ if they can exchange information with each other. 
These neighbor relationships can be described by graphs. For each index $i\in\{1,\ldots,m\}$, define a row-neighbor graph $\bbb G_i^{{\rm row}}$ with $m$ vertices to represent the neighbor relationships among agents $\dbb{ij}$, $j\in\{1,\ldots,m\}$ in that $(k,\ell)$ is an edge in $\bbb G_i^{{\rm row}}$ whenever agents $\dbb{ik}$ and $\dbb{i\ell}$ are row-neighbors. Similarly, define a column-neighbor graph $\bbb G_i^{{\rm col}}$ with $m$ vertices to represent the neighbor relationships among agents $\dbb{ij}$, $i\in\{1,\ldots,m\}$ in that $(k,\ell)$ is an edge in $\bbb G_i^{{\rm col}}$ whenever agents $\dbb{kj}$ and $\dbb{\ell j}$ are column-neighbors.
For simplicity, the following homogeneous neighbor graph assumption is adopted throughout the paper. 
Our algorithm readily extends to heterogeneous neighbor graph settings, albeit with more involved algorithmic expressions.

\vspace{.05in}

\begin{assumption}
    All $m$ row-neighbor graphs $\bbb G_i^{{\rm row}}$, $i\in\{1,$ $\ldots,m\}$ are identical and connected. All $m$ column-neighbor graphs $\bbb G_i^{{\rm col}}$, $i\in\{1,\ldots,m\}$ are identical and~connected. 
\end{assumption}

\vspace{.05in}

Under the above assumption, we use $\bbb G^{{\rm row}}$ and $\bbb G^{{\rm col}}$ to denote the common row- and column-neighbor graphs, respectively, both with vertex set $\{1,\ldots,m\}$ and connected.
Let $\scr N_{ij}$ and $\scr M_{ij}$ denote the row- and column-neighbor sets of agent $\dbb{ij}$, respectively. 
For notational consistency, we impose $\scr N_{ij},\scr M_{ij}\subseteq\{1,\ldots,m\}$. Under this convention, $k\in\scr N_{ij}$ indicates that agent $\dbb{ik}$ is a row-neighbor of agent $\dbb{ij}$, and similarly, $k\in\scr M_{ij}$ means that agent $\dbb{kj}$ is a column-neighbor of agent $\dbb{ij}$.
For presentation simplicity, we also assume that each agent is always a row- and column-neighbor of itself, so $j\in\scr{N}_{ij}$ and $i\in\scr{M}_{ij}$ for all $i,j$.


\subsection{Distributed Optimization Reformulation}

We first reformulate the distributed quantum least-squares problem described in the previous subsection as a distributed optimization problem, since VQLS partially relies on a classical optimizer and our algorithm will leverage ideas from distributed optimization.

To this end, let $L$ denote the Laplacian matrix of the row neighbor graph $\bbb G^{{\rm row}}$, and define $\bar L = L\otimes I$, where $I$ is the $2^q\times 2^q$ identity matrix and $\otimes$ denotes the Kronecker product. 
Let $A_i$ denote the entire $i$th row block of $A$, i.e., 
$A_i = [A_{i1},\cdots,A_{im}]$. Then, $\|Ax-b\|^2=\sum_{i=1}^m\|A_i x-b_i\|^2$.
Define $\bar A_i$ as the block diagonal matrix whose $j$th diagonal block is $A_{ij}$, and define $\bar b_i$ as the stacked vector formed by $b_{ij}$; that is, 
$\bar A_i = \diag (A_{i1}, \ldots, A_{im})$ 
and $ \bar b_i = [b_{i1}^\top, \cdots, b_{im}^\top]^\top$. 
With these notations, we have the following lemma, which provides an alternative equivalent expression for the least squares solutions to $Ax=b$.

\vspace{.05in}

\begin{lemma}\label{lemma:goal_fixed}
If $\bbb G^{{\rm row}}$ and $\bbb G^{{\rm col}}$ are connected, then $x^*$ is a least squares solution to $Ax=b$ if, and only if, there exist $\{z_i^*\}_{i=1}^m$ such that $(x^*,\{z_i^*\}_{i=1}^m)$ is a minimizer of $\sum_{i=1}^m \|\bar A_i x-\bar b_i-\bar Lz_i\|^2$.
\end{lemma}

\vspace{.05in}

The proof of the lemma can be found in Appendix A.
The lemma is motivated by \cite[Lemma 2.1]{cao2017continuous}. A similar result is derived in \cite[Lemma 3.1]{huang2022scalable} via a different proof technique.

Although the equivalent optimization problem in Lemma~\ref{lemma:goal_fixed} appears more complex and involves more variables, it is well suited to our block-partitioned distributed setting for the following reason. 
Note that $x$ and each $z_i$ are vectors of dimension $2^qm$. Uniformly partition them into $m$ subvectors, i.e., $x = [x_{1}^\top, \cdots, x_{m}^\top]^\top$ and $z_i = [z_{i1}^\top, \cdots, z_{im}^\top]^\top$, with all $x_k$ and $z_{ik}$ of dimension $2^q$. Then,
\begin{align}
    & \textstyle\sum_{i=1}^m\|\bar A_ix-\bar b_i-\bar Lz_i\|^2 \nonumber\\
    =\; & \textstyle\sum_{i=1}^m \sum_{j=1}^m \| A_{ij} x_{j} - b_{ij} - \sum_{k \in \mathcal{N}_{ij}} (z_{ij} - z_{ik}) \|^2. \label{eq:distopt}
\end{align}
Let $f_{ij} = \| A_{ij} x_{j} - b_{ij} - \sum_{k \in \mathcal{N}_{ij}} (z_{ij} - z_{ik}) \|^2$ and treat it as the local cost function associated with agent $\dbb{ij}$. 
From \eqref{eq:distopt} and Lemma \ref{lemma:goal_fixed}, finding a least squares solution to $Ax=b$ is equivalent to a distributed optimization problem, whose objective $f = \sum_{i=1}^m \sum_{j=1}^mf_{ij}$ decomposes into a sum of local cost functions over the $m^2$ agents.
To solve this distributed optimization problem, each agent $\dbb{ij}$ has control over two variables, $x_{ij}$ and $z_{ij}$, which serve as estimates of $x_j^*$ and $z_{ij}^*$, respectively. 
The two variables will be prepared as quantum states and measured on the agent's quantum computer.
Note that the cost function $f_{ij}$ depends only on the local information available to agent $\dbb{ij}$ and information from its row-neighbors, and is therefore locally computable.

It is worth noting that the above distributed optimization formulation differs from standard ones due to the block partition structure, which induces two distinct coupling mechanisms. Each agent interacts with both row-neighbors and column-neighbors.
Column-neighbors enforce consensus on the corresponding component of a least squares solution by exchanging the variables $x_{ij}$ (with common $j$) within each column. 
Row-neighbors, on the other hand, coordinate through the variables $z_{ij}$ to ensure that the collection of their variables $x_{ij}$ (with common $i$) jointly satisfies the least squares objective.

It is also worth emphasizing that, although the distributed optimization formulation in \eqref{eq:distopt} admits a discrete-time distributed (classical) gradient descent implementation, such an approach is not directly compatible with quantum computing. In Subsection \ref{subsec:gradient_descent_fail}, we will demonstrate that applying a distributed gradient descent scheme to the quantum setting fails to yield a valid algorithm.
Moreover, certain coordination and computation terms 
in classical distributed optimization cannot be directly evaluated within quantum circuits. These fundamental limitations complicate the design of distributed quantum algorithms, as elaborated in the next section.

\subsection{The Distributed VQLS Algorithm}\label{subsec:algorithm}

This subsection presents the proposed distributed variational quantum linear equation solver, which consists of four main components: data encoding, variational ansatz, cost function, and optimization procedure at each agent.
The first two components define the quantum state representation, the cost function specifies the optimization objective (whose value is estimated using quantum circuits), and the last corresponds to the classical optimization process; together, they form a hybrid quantum-classical framework. 
We now describe each component in detail; all components are identical across agents.



{\bf Data Encoding:}
Agent $\dbb{ij}$ represents $A_{ij}$ using a linear combination of unitaries, that is, $A_{ij}=\sum_{k} c_{ijk} A_{ijk}$, where $k$ is a positive integer, $c_{ijk}$ are coefficients, and $A_{ijk}$ are unitary matrices. 
Each unitary $A_{ijk}$ is implemented as a quantum circuit. The vector $b_{ij}$ is encoded as a quantum state $\ket{b_{ij}} = U_{ij} \ket{0}$, where $U_{ij}$ is a state preparation unitary encoding the entries of $b_{ij}$ into quantum amplitudes, while its norm $\|b_{ij}\|$ handled separately; in other words, 
$b_{ij}=\|b_{ij}\|\ket{b_{ij}}=\|b_{ij}\|U_{ij}\ket{0}$.

As mentioned earlier, constructing an LCU representation for a matrix is in general nontrivial, especially for large size matrices. 
In the proposed distributed algorithm, the LCU construction is performed locally at each agent on smaller sized matrices corresponding to individual blocks. This reduces the computational burden. 

\vspace{.05in}

{\bf Variational Ansatz:}
Agent $\dbb{ij}$ prepares two quantum states $\ket{x_{ij}}$ and $\ket{z_{ij}}$ using parameterized quantum circuit architectures with parameter vectors $\alpha_{ij}$ and $\beta_{ij}$, respectively:
\begin{align*}
    & \ket{x_{ij}(\alpha_{ij})} = U(\alpha_{ij})\ket{0}, 
    & U(\alpha_{ij})=\textstyle\prod_{k=1}^{p} G_k(\alpha_{ijk}), \\
    & \ket{z_{ij}(\beta_{ij})} = V(\beta_{ij})\ket{0}, 
    & V(\beta_{ij})=\textstyle\prod_{k=1}^{q} H_k(\beta_{ijk}),
\end{align*}
where $V(\cdot)$ and $U(\cdot)$ denote fixed structure parameterized quantum circuits, $p$ and $q$ are the corresponding circuit depths, and $\{G_k\}_{k=1}^p$ and $\{H_k\}_{k=1}^q$ are predefined sequences of parameterized quantum gates. The parameters $\alpha_{ij}=(\alpha_{ij1},\ldots,\alpha_{ijp})$ and $\beta_{ij}=(\beta_{ij1},\ldots,\beta_{ijq})$ are trainable real-valued vectors.

Here we adopt a fixed structure ansatz, in which the circuit layout is predefined and only the parameters are optimized during training. Specifically, the hardware efficient ansatz consists of low-depth single-qubit rotations and nearest-neighbor entangling gates. 
In the numerical experiments in Section \ref{sec:simulation}, it is restricted to $R_y$ rotations and nearest-neighbor controlled-$Z$ gates, which ensures that the resulting quantum states remain real-valued.
This design yields shallow circuits compatible with realistic hardware connectivity, while retaining sufficient expressivity for the problems considered in this paper.


\vspace{.05in}

\begin{figure*}
\hrule
\begin{align}
\braket{s_{ij}}{s_{ij}} =\;& \rho_{ij}^2 \bra{x_{ij}} A_{ij}^\dagger A_{ij} \ket{x_{ij}} 
- 2 \rho_{ij} \textstyle\sum_{k \in \mathcal{N}_{ij}} \big( \sigma_{ij} {{\rm Re}}\bra{x_{ij}} A_{ij}^\dagger \ket{z_{ij}} - \sigma_{ik} {{\rm Re}}\bra{x_{ij}} A_{ij}^\dagger \ket{z_{ik}} \big)
\nonumber \\
&\! + \textstyle\sum_{k \in \mathcal{N}_{ij}} \sum_{\ell \in \mathcal{N}_{ij}} \big( \sigma_{ij}^2 \braket{z_{ij}}{z_{ij}} - \sigma_{ij}\sigma_{i\ell} \braket{z_{ij}}{z_{i\ell}} - \sigma_{ij}\sigma_{ik} \braket{z_{ik}}{z_{ij}} + \sigma_{ik}\sigma_{i\ell} \braket{z_{ik}}{z_{i\ell}} \big), \label{eq:ss} \\
\braket{b_{ij}}{s_{ij}} =\;& \rho_{ij} \bra{b_{ij}} A_{ij} \ket{x_{ij}} - \textstyle\sum_{k \in \mathcal{N}_{ij}} \big( \sigma_{ij} \braket{b_{ij}}{z_{ij}} - \sigma_{ik} \braket{b_{ij}}{z_{ik}} \big). \label{eq:bs}
\end{align}
\hrule
\end{figure*}


{\bf Cost Function:}
Agent $\dbb{ij}$ maintains a local cost function 
$C_{ij} = \| A_{ij} x_{ij} - b_{ij} - \sum_{k \in \mathcal{N}_{ij}} (z_{ij} - z_{ik}) \|^2$
and evaluates it independently by estimating quantum expectation values.
To enable quantum evaluation, each classical vector in the cost function needs to be encoded into a normalized quantum state, with its norm treated separately. In particular, let
$$x_{ij}=\rho_{ij}\ket{x_{ij}} \;\;\;{\rm and} \;\;\; z_{ij}=\sigma_{ij}\ket{z_{ij}},$$
where $\rho_{ij}, \sigma_{ij} >0$ denote local trainable parameters corresponding to the vector norms. 
With these representations, 
\[
C_{ij}
= \big\| \rho_{ij}A_{ij}\ket{x_{ij}} - b_{ij} - \sum_{k \in \mathcal{N}_{ij}} \big( \sigma_{ij}\ket{z_{ij}} - \sigma_{ik}\ket{z_{ik}} \big) \big\|^2.
\]
To simplify the expressions, define the intermediate quantity
$$
s_{ij} = \rho_{ij}A_{ij}\ket{x_{ij}} - \textstyle\sum_{k \in \mathcal{N}_{ij}} ( \sigma_{ij}\ket{z_{ij}} - \sigma_{ik}\ket{z_{ik}} ).    
$$
Then, the local objective function can be expressed as
$$C_{ij} = \|s_{ij} - b_{ij} \|^2 = \braket{s_{ij}}{s_{ij}} + \braket{b_{ij}}{b_{ij}} - 2 {{\rm Re}}(\braket{b_{ij}}{s_{ij}}).$$
Recall that $b_{ij}=\|b_{ij}\|U_{ij}\ket{0}$, so $\braket{b_{ij}}{b_{ij}} = \|b_{ij}\|^2$. 
For the other two inner product terms $\braket{s_{ij}}{s_{ij}}$ and $\braket{b_{ij}}{s_{ij}}$, it is straightforward to verify that they can be written as \eqref{eq:ss} and \eqref{eq:bs}, respectively, where $\dagger$ denotes the Hermitian adjoint (conjugate transpose).
The quantum terms in these two equations can be implemented using the state preparation circuit $U_{ij}$ and the parameterized variational circuits $U(\alpha_{ij})$ and $V(\beta_{ij})$, as described below: 
\begin{align} 
\bra{x_{ij}} A_{ij}^\dagger A_{ij} \ket{x_{ij}} &= \bra{0} U^\dagger(\alpha_{ij}) A_{ij}^\dagger A_{ij} U(\alpha_{ij}) \ket{0}, \label{eq:overlap_test} \\
\bra{x_{ij}} A_{ij}^\dagger \ket{z_{ik}} &= \bra{0} U^\dagger(\alpha_{ij}) A_{ij}^\dagger V(\beta_{ik}) \ket{0}, \label{eq:hadamard1} \\
\braket{z_{ik}}{z_{i\ell}} &= \bra{0} V^\dagger(\beta_{ik}) V(\beta_{i\ell}) \ket{0}, \label{eq:hadamard2} \\
\bra{b_{ij}} A_{ij} \ket{x_{ij}} &= \bra{0} U_{ij}^\dagger A_{ij} U(\alpha_{ij}) \ket{0}, \label{eq:hadamard3} \\ 
\braket{b_{ij}}{z_{ik}} &= \bra{0} U_{ij}^\dagger V(\beta_{ik}) \ket{0}. \label{eq:hadamard4}
\end{align}
These inner product terms are estimated via the standard Hadamard test. 
The corresponding quantum circuits are given in Appendix~B.
With \eqref{eq:overlap_test}--\eqref{eq:hadamard4} and their quantum circuits, agent $\dbb{ij}$ is able to estimate its cost function $C_{ij}$ via local quantum measurements.
The intuition behind the design of $C_{ij}$ has been explained in \eqref{eq:distopt} and the subsequent paragraph.

In the proposed distributed VQLS algorithm, agent $\dbb{ij}$ also needs to estimate the gradients of $C_{ij}$ with respect to the parameter vectors $\alpha_{ij}$ and $\beta_{ij}$. These gradients can be straightforwardly evaluated using the parameter shift rule \cite{schuld2019evaluating}. Key explicit expressions are deferred to Appendix C, and detailed derivation procedure can be found in \cite[Appendix F]{vqls}.


\vspace{.05in}

{\bf Optimization Procedure:}
Agent $\dbb{ij}$ performs an iterative optimization procedure on a local classical processor, with cost function values and gradients estimated via quantum circuit evaluations on its local quantum processor, thereby forming a hybrid quantum-classical optimization loop.
The classical optimization iteration employs ideas from the DIGing algorithm \cite{34_nedic2017diging} and the Adam optimizer~\cite{41_kingma2015adam}. 
The DIGing algorithm is a distributed gradient tracking method for consensus optimization over networks, and its update mechanism is applied among column-neighboring agents (i.e., agents $\dbb{ij}$ with fixed $j$).
The Adam optimizer is an adaptive gradient method that computes parameter updates based on estimates of the first and second moments of the gradients, and these updates are applied across both row-neighboring and column-neighboring agents.

Recall that the optimization variables $x_{ij}$ and $z_{ij}$ in the local cost function $C_{ij}$ at agent $\dbb{ij}$ are parameterized by $(\rho_{ij}, \alpha_{ij})$ and $(\sigma_{ij}, \beta_{ij})$, respectively, where scalars $\rho_{ij}, \sigma_{ij}$ are norm parameters and vectors $\alpha_{ij}, \beta_{ij}$ are quantum circuit parameters. At each iteration, the quantum processor evaluates $C_{ij}$ and its gradients with respect to $\rho_{ij}, \alpha_{ij}, \sigma_{ij}, \beta_{ij}$; the classical optimizer then updates these parameters using the estimated gradients and information received from neighbors. 
Since $\rho_{ij}$ and $\alpha_{ij}$ will be updated in the same manner, and the same holds for $\sigma_{ij}$ and $\beta_{ij}$, we define the following stacked parameter vectors for algorithmic simplicity: 
$$\tilde{\alpha}_{ij} = [\alpha_{ij}^\top,\rho_{ij}]^\top \;\;\; {\rm and} \;\;\; 
\tilde{\beta}_{ij} = [\beta_{ij}^\top,\sigma_{ij}]^\top.$$
In addition to $\tilde{\alpha}_{ij}$ and $\tilde{\beta}_{ij}$, agent $\dbb{ij}$ also has control over auxiliary variables for both optimization and coordination. Specifically, $\mu_{ij}$ and $\nu_{ij}$ are the first- and second-moment estimates used in the Adam updates for $\tilde{\alpha}_{ij}$ (with column-neighbors), $\mu'_{ij}$ and $\nu'_{ij}$ are the corresponding moment estimates for $\tilde{\beta}_{ij}$ (with row-neighbors), $\eta_{ij}$ is the adaptive stepsize shared by both Adam updates, and $y_{ij}$ is the auxiliary variable for gradient tracking (with column-neighbors).

With the above description and notation, the proposed distributed VQLS algorithm is presented as follows.

\vspace{.05in}

\noindent
{\bf Distributed VQLS Algorithm} (implemented at each agent)

\vspace{.05in}

\noindent
{\bf Initialization:}
At time $t=0$, agent $\dbb{ij}$ sets $\mu_{ij}(0)$, $\nu_{ij}(0)$, $\mu'_{ij}(0)$, $\nu'_{ij}(0)$ to zero vectors, 
randomly initializes $\tilde\alpha_{ij}(0)$ and $\tilde\beta_{ij}(0)$, 
transmits $\tilde\beta_{ij}(0)$ to its row-neighbors (concurrently, it receives the corresponding quantities from its row-neighbors), sets
$y_{ij}(0) = \nabla_{\tilde\alpha_{ij}}C_{ij}(0)$ and a dummy previous gradient $\nabla_{\tilde\alpha_{ij}}C_{ij}(-1) = \nabla_{\tilde\alpha_{ij}}C_{ij}(0)$, 
where $\nabla_{\tilde\alpha_{ij}}C_{ij}(0)$ is estimated on the local quantum processor.

\vspace{.05in}

\noindent
\textbf{Iteration:}
Between clock times $t$ and $t+1$, agent $\dbb{ij}$ performs the steps enumerated below in the order indicated.

\begin{enumerate}

\item \textbf{1st Transmission:}
Agent $\dbb{ij}$ transmits $\tilde\alpha_{ij}(t)$ and $y_{ij}(t)$ to its column-neighbors and $\tilde\beta_{ij}(t)$ to its row-neighbors; simultaneously, it receives the corresponding quantities from its row- and column-neighbors. 

\vspace{.03in}

\item \textbf{1st Update:}
Agent $\dbb{ij}$ first computes 
\begin{align}
    \mu_{ij}(t+1)
    &= \gamma_1 \mu_{ij}(t)
    + (1-\gamma_1) y_{ij}(t), \label{eq:alpha_Adam1} \\
    \nu_{ij}(t+1)
    &= \gamma_2 \nu_{ij}(t)
    + (1-\gamma_2) (y_{ij}(t))^{\odot 2}, \label{eq:alpha_Adam2} 
\end{align}
where $\gamma_1,\gamma_2$ are the exponential decay rates of the Adam optimizer, and $\odot$ denotes the element-wise product, so that $(\cdot)^{\odot 2}$ represents the element-wise square.\footnote{
For a vector $x$, $x^{\odot 2} = x \odot x$, i.e., $(x^{\odot 2})_k = x_k^2$ for all $k$.
}
Agent $\dbb{ij}$ next estimates $\nabla_{\tilde\alpha_{ij}}C_{ij}(t)$ and $\nabla_{\tilde\beta_{ik}}C_{ij}(t)$, $k\in\scr N_{ij}$ on the quantum processor, and then computes
\begin{align}
    \tilde\alpha_{ij}(t+1)
    = &
    \sum_{k\in\mathcal M_{ij}} w_{ik}\tilde\alpha_{kj}(t) \nonumber \\
    &\;\; -
    \eta(t+1)
    \frac{\mu_{ij}(t+1)}
    {\sqrt{\nu_{ij}(t+1)}+\epsilon}, \label{eq:alpha_update} 
\end{align}
\begin{align}
    y_{ij}(t+1)
    = &
    \sum_{k\in\mathcal M_{ij}} w_{ik}y_{kj}(t) \nonumber \\
    &\;\; +
    \nabla_{\tilde\alpha_{ij}}C_{ij}(t)
    -
    \nabla_{\tilde\alpha_{ij}}C_{ij}(t-1), \label{eq:tracker_y_update}   
\end{align}
where 
$    \eta(t+1)
    =
    \eta \sqrt{1-\gamma_2^{t+1}}/(1-\gamma_1^{t+1})
    $
with $\eta>0$ being the base stepsize, 
the division and square root in \eqref{eq:alpha_update} are performed element-wise,\footnote{
For vectors $x,z\in\R^d$, $\frac{x}{\sqrt{z} + \epsilon}
=
\big[
\frac{x_{1}}{\sqrt{z_{1}}+\epsilon}, \cdots,
\frac{x_{d}}{\sqrt{z_{d}}+\epsilon}
\big]^\top.
$
} 
$\epsilon>0$ is a small constant for numerical stability (avoids division by zero),
and 
$w_{ik}$ are the Metropolis weights defined as\footnote{
The weights follow the Metropolis algorithm in \cite{metro2}, originally proposed for solving the distributed averaging problem over undirected graphs.
} 
$$
w_{ik}=
\begin{dcases}
\displaystyle\frac{1}{\max\{ |\mathcal{M}_{ij}|, |\mathcal{M}_{kj}| \}}, & k \in \mathcal{M}_{ij} \setminus \{i\}, \\[0.5ex]
1 - \displaystyle\sum_{\ell \in \mathcal{M}_{ij} \setminus \{i\}} w_{i\ell},  & k=i.
\end{dcases}
$$


\item \textbf{2nd Transmission:}
Agent $\dbb{ij}$ transmits $\nabla_{\tilde\beta_{ik}}C_{ij}(t)$, $k\in\scr N_{ij}$ to its row-neighbors; concurrently, it receives the corresponding quantities from its row-neighbors.

\vspace{.03in}

\item \textbf{2nd Update:}
Agent $\dbb{ij}$ computes  
\begin{align}
    \mu'_{ij}(t+1)
    &= \gamma_1 \mu'_{ij}(t)
    + (1-\gamma_1) g_{ij}(t),         \label{eq:beta_Adam1} \\[.65ex] 
    \nu'_{ij}(t+1)
    &= \gamma_2 \nu'_{ij}(t)
    + (1-\gamma_2) (g_{ij}(t))^{\odot 2}, \label{eq:beta_Adam2} \\
    \tilde\beta_{ij}(t+1)
    &=
    \tilde\beta_{ij}(t)
    -
    \eta(t+1)
    \frac{\mu'_{ij}(t+1)}
    {\sqrt{\nu'_{ij}(t+1)}+\epsilon},\!\! \label{eq:beta_update}     
\end{align}
where $g_{ij}(t) = \sum_{k\in\scr N_{ij}}\nabla_{\tilde\beta_{ij}}C_{ik}(t)$.

\end{enumerate}

\vspace{.1in}

It is straightforward to check that the implementation of all four components requires only local information at each agent and information from its neighbors. Therefore, the proposed algorithm is fully distributed.

For a concise presentation of the algorithm, we refer to the pseudocode {\bf Algorithm 1: Distributed VQLS}.


\begin{algorithm}[t]
\LinesNumbered
\caption{\textbf{Distributed VQLS (implemented at each agent)}}
\label{alg:distributed-vqls}
\setcounter{AlgoLine}{0}
\DontPrintSemicolon
\KwIn{ $A_{ij}, b_{ij}$, $\scr N_{ij}, \scr M_{ij}$;

hyperparameters: $\gamma_1,\gamma_2,\eta,\epsilon,\varepsilon_{\rm stop},T$}
\textbf{Initialization:} $\mu_{ij}(0)=\nu_{ij}(0)=\mu'_{ij}(0)=\nu'_{ij}(0)=0$; randomly initialize $\tilde{\alpha}_{ij}(0)$ and $\tilde{\beta}_{ij}(0)$; send $\tilde{\beta}_{ij}(0)$ to row-neighbors and receive the corresponding values;  $y_{ij}(0)=\nabla_{\tilde{\alpha}_{ij}}C_{ij}(0)$; 
$\nabla_{\tilde{\alpha}_{ij}}C_{ij}(-1)=\nabla_{\tilde{\alpha}_{ij}}C_{ij}(0)$\;
\BlankLine
\For{$t=0,\ldots,T$}{
Send $\tilde{\alpha}_{ij}(t)$ and $y_{ij}(t)$ to column-neighbors, send $\tilde{\beta}_{ij}(t)$ to row-neighbors; receive the corresponding quantities\; \hfill \tcp*{1st transmission}
Update $\mu_{ij}(t+1)$ and $\nu_{ij}(t+1)$ using \eqref{eq:alpha_Adam1}--\eqref{eq:alpha_Adam2}\;
$\tilde{\alpha}_{ij}(t+1)=\sum_{k\in\mathcal M_{ij}} w_{ik}\tilde{\alpha}_{kj}(t)-\eta(t+1)\frac{\mu_{ij}(t+1)}{\sqrt{\nu_{ij}(t+1)}+\epsilon}$\;
$y_{ij}(t+1)=\sum_{k\in\mathcal M_{ij}} w_{ik}y_{kj}(t)+\nabla_{\tilde{\alpha}_{ij}}C_{ij}(t)-\nabla_{\tilde{\alpha}_{ij}}C_{ij}(t-1)$\;\tcp*[r]{1st update}
Send $\{\nabla_{\tilde{\beta}_{ik}}C_{ij}(t)\}_{k\in\scr N_{ij}}$ to row-neighbors and receive the corresponding values\;\tcp*[r]{2nd transmission}
Update $\mu'_{ij}(t+1)$ and $\nu'_{ij}(t+1)$ using \eqref{eq:beta_Adam1}--\eqref{eq:beta_Adam2}\;
$\tilde{\beta}_{ij}(t+1)=\tilde{\beta}_{ij}(t)-\eta(t+1)\frac{\mu'_{ij}(t+1)}{\sqrt{\nu'_{ij}(t+1)}+\epsilon}$\;\tcp*[r]{2nd update}
\If{$\|Ax-b\|\le \varepsilon_{\rm stop}$}{
\textbf{break}\;\tcp*[r]{termination criterion}
}
}
\textbf{end}\;
\end{algorithm}


We provide the following discussion to facilitate understanding of the proposed algorithm.

The updates in \eqref{eq:beta_Adam1}--\eqref{eq:beta_update} implement collective Adam optimization among row-neighboring agents, while those in \eqref{eq:alpha_Adam1}--\eqref{eq:tracker_y_update} implement a delayed gradient-tracking DIGing scheme with Adam among column-neighboring agents. 
Distributed gradient tracking enables agents to asymptotically track the network-average gradient via consensus, while Adam provides adaptive momentum for smoother updates; together, they yield faster and more stable convergence. 
The idea of incorporating Adam into distributed gradient tracking was studied in \cite{gtadam} for classical distributed online (convex) optimization to achieve faster convergence.



The proposed algorithm uses quantum circuits to evaluate gradients with respect to quantum circuit parameters. If the gradients with respect to the classical variables $x_{ij}$ and $z_{ij}$ are instead computed on local classical processors, the algorithm reduces to a classical counterpart.
Since the dimensions of the quantum circuit parameters are much smaller than those of $x_{ij}$ and $z_{ij}$, the proposed quantum distributed algorithm reduces communication costs compared with its classical counterpart.

When implementing the classical counterpart, each agent needs only one transmission round with its neighbors. In contrast, the proposed quantum distributed algorithm requires two communication rounds, since certain terms needed for classical gradient computation cannot be directly evaluated on quantum processors; this induces a delayed gradient-tracking variant of the DIGing scheme in \eqref{eq:tracker_y_update}.

Both gradient tracking and Adam can be viewed as first-order optimization methods that reduce to gradient descent under simplifications. Such a simplified distributed scheme is effective in classical settings, since the least squares linear equation problem is (strongly) convex. However, numerical experiments in Subsection \ref{subsec:gradient_descent_fail} show that the corresponding quantum distributed algorithm does not always function; this may be due to the highly nonlinear and nonconvex nature of quantum circuit evaluations.

In the proposed distributed VQLS algorithm, the communication overhead per iteration scales with the number of variational parameters rather than with the (exponentially large) dimension of the local solution vectors. For instance, for a linear system of dimension \(2^{51}\) to be solved over a four-agent network (see Subsection \ref{subsec:51} for a concrete example), each agent operates on a 50-qubit local variational state represented by \(3\times 50=150\) rotation parameters together with an additional normalization factor. The four communicated quantities are therefore represented by only 151 real entries, whereas explicitly representing and transmitting a local state vector would require \(2^{50}\) entries. 
Such exponential storage and communication requirements make explicit classical vector-based computation and inter-agent communication infeasible at this scale. By representing the local solutions implicitly as parameterized quantum states and exchanging only compact variational parameters, the distributed VQLS enables linear systems of dimension \(2^{51}\) to be addressed while maintaining feasible communication among agents.



\section{Numerical Experiments}\label{sec:simulation}

This section presents simulation results that validate the proposed distributed algorithm and evaluate its performance under various scenarios.

We consider linear systems derived from Ising-inspired Hamiltonians (of the transverse field type) \cite{PFEUTY197079}
and cluster-state stabilizers (with perturbations) \cite{gottesman}
as structured test instances. 
Ising-inspired Hamiltonians have been commonly used in prior quantum algorithms such as VQLS \cite{vqls}, while cluster-state stabilizer systems provide additional structured instances relevant to quantum many-body settings.
Compared with generic linear systems, these structured instances enable us to evaluate the proposed distributed algorithm on quantum-native problem classes where the system matrices admit efficient LCU decompositions.
All simulation results, shown as dotted lines, are averaged over 30 independent trials 
(corresponding to different random initializations)
to account for the stochastic nature of a quantum algorithm. The shaded region represents error bars corresponding to one standard deviation across all runs. 

\subsection{Large-Scale Demonstration}\label{subsec:51}

We first demonstrate the performance of the proposed algorithm on a large-scale instance to illustrate its scalability.

In the VQLS work \cite{vqls}, simulations of Ising-type linear systems are demonstrated at scales up to 50 qubits (matrix $A$ of dimension $2^{50} \times 2^{50}$), which represents the largest scale considered therein. 
Following this setting, we consider a 51-qubit (matrix $A$ of dimension $2^{51} \times 2^{51}$) instance to assess scalability.
Specifically, the structured linear system $Ax=b$ defined on $n=51$ qubits is given by
\begin{align} 
    A &= \textstyle\frac{1}{\zeta} \big( \sum_{i=1}^{n} X_i + \kappa \sum_{i=1}^{n-1} Z_i Z_{i+1} + \lambda I \big), \label{eq:Ising_A}\\
    b & = H^{\otimes n} \ket{0}^{\otimes n}, \label{eq:Ising_b}
\end{align}
where $X_i$ and $Z_i$ denote the Pauli-$X$ and Pauli-$Z$ operators acting on the $i$th qubit, $I$ is the identity matrix of appropriate dimension, $\zeta$ is a positive normalization constant, $\kappa$ and $\lambda$ are real scalars, $H$ denotes the Hadamard gate, $(\cdot)^{\otimes n}$ denotes the $n$-fold tensor product, 
and $\ket{0}^{\otimes n}$ denotes the $n$-qubit initial state in the computational basis.
Here, the system matrix $A$ is constructed from an Ising-inspired Hamiltonian of the transverse field type, and the vector $b$ is prepared as a uniform superposition state. In particular, the term $Z_i Z_{i+1}$ represents a nearest-neighbor interaction between qubits $i$ and $i+1$, while $\sum_{i=1}^n X_i$ corresponds to a transverse field term acting on each qubit; the scalar $\kappa$ controls the interaction strength, and $\lambda$ 
ensures desired spectral properties (e.g., positive definiteness). 
Such a matrix $A$ admits a decomposition into a linear combination of Pauli operators, making it well suited for quantum circuit evaluation.

We partition $A$ into $2 \times 2$ blocks and accordingly employ a 4-agent network to solve $Ax = b$, with each agent handling a 50-qubit subsystem. 
With a small coupling parameter $\kappa$, the solution state $\ket{x}$ and its partition $\ket{x_i}$ admit simple matrix product state (MPS) representations 
due to their low entanglement structure, enabling efficient classical simulation for the $n=51$ instance.
In the simulations, we set the coupling parameter to $\kappa = 0.1$ and select the normalization constant $\zeta$ and shift parameter $\lambda$ such that the condition number of $A$ is $50$. For each agent, the ansatz consists of one layer of Hadamard gates acting on all qubits, followed by three alternating layers of single-qubit $R_y$ rotations on each qubit and controlled-$Z$ gates acting on each nearest-neighbor qubit pair. The parameters in $\alpha_{ij}$ and $\beta_{ij}$ are randomly initialized in $[-\frac{\pi}{6}, \frac{\pi}{6}]$, while $\rho_{ij}$ and $\sigma_{ij}$ are initialized to $1$, and results are averaged over $10$ random initializations with the base stepsize $\eta=0.02$. 

\begin{figure}[!t]
    \centering
    \includegraphics[width=0.485\textwidth]{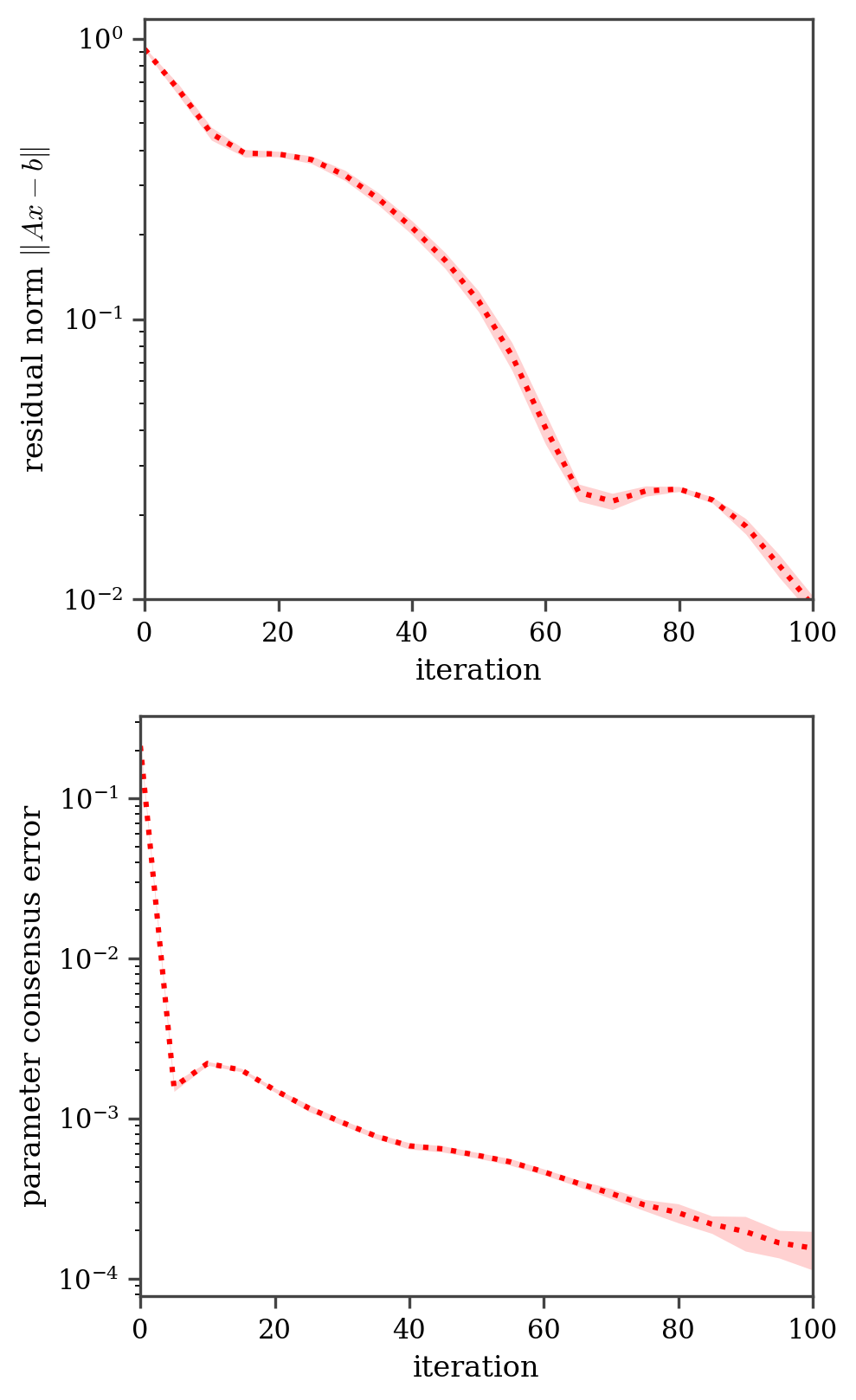}
    \caption{Global residual norm and parameter consensus error trajectories for the proposed distributed VQLS algorithm applied to the 51-qubit Ising-inspired linear system in \eqref{eq:Ising_A} and \eqref{eq:Ising_b} with a $2 \times 2$ block partition and a 4-agent network}
    \label{fig:50qubits}
\end{figure}

Since the full solution vector lies in a $2^{51}$-dimensional Hilbert space, directly reconstructing the simulated solution estimate $x$ is computationally expensive.
To get around this, we evaluate the global residual $\|Ax - b\|$ via
$\|Ax - b\| = \sqrt{
\| r_1 \|^2 + \| r_2 \|^2}$, where $r_1 = A_{11} \bar{x}_1 + A_{12} \bar{x}_2 - b_1$ and 
$r_2 = A_{21} \bar{x}_1 + A_{22} \bar{x}_2 - b_2$, with $\bar{x}_1 = \frac{1}{2}(\rho_{11} \ket{x_{11}} + \rho_{21} \ket{x_{21}})$
and $\bar{x}_2 = \frac{1}{2}(\rho_{12} \ket{x_{12}} + \rho_{22} \ket{x_{22}})$. 
The residual norm can thus be computed via overlap evaluations such as $\rho_{11}^2 \bra{x_{11}} A_{i1}^\dagger A_{i1} \ket{x_{11}}$, which are directly accessible from the MPS representation. Meanwhile, we report the parameter consensus error, defined as the standard deviation among the variational parameters $\tilde\alpha_{ij}$ across different rows, as a proxy for the agreement level among agents.
The trajectories of the global residual norm and the parameter consensus error are plotted in Figure \ref{fig:50qubits}.
As shown in the figure, the residual norm decreases steadily over 100 iterations, dropping by nearly two orders of magnitude. 
Meanwhile, the parameter consensus error decreases rapidly in the initial stage and continues to decline to the $10^{-4}$ level. 
These results indicate that the proposed distributed VQLS algorithm remains stable and is capable of solving linear systems of size $2^{51} \times 2^{51}$.

This experiment demonstrates that a 51-qubit problem can be solved using multiple 50-qubit processors, highlighting the ability of the proposed framework to solve a larger global system using only smaller quantum devices. 
This is particularly relevant given current hardware limitations, where large-scale, high-fidelity quantum processors remain unavailable. 
By distributing the information and computation across a multi-agent network, the proposed approach effectively bypasses these limitations and provides a scalable pathway for solving higher-dimensional quantum linear systems.

\subsection{Effect of Block Partitioning}

We next consider a smaller problem size to enable the use of Hadamard test circuits for estimating overlaps and expectation values required in the cost functions and their gradients. 
Meanwhile, a large matrix $A$ can be partitioned into square blocks in different ways, where a larger number of blocks corresponds to smaller block sizes. We therefore investigate the effect of different block partitioning strategies on the performance of the proposed distributed algorithm.

To this end, we consider a 13-qubit linear system, where the system matrix $A$ has size $2^{13} \times 2^{13}$, and partition it in three ways: $2 \times 2$, $4 \times 4$, and $8 \times 8$ blocks, where each block corresponds to subsystems of size $2^{12}\times 2^{12}$, $2^{11}\times 2^{11}$, and $2^{10}\times 2^{10}$, respectively.
To control the wall-clock running time for finer block partitions, we adopt a perturbed cluster-state stabilizer system, in which the system matrix $A$ consists of six Pauli terms, expressed as
\begin{equation*}
A\!=\!c_1 I +c_2[X_1Z_2+Z_3X_4Z_5+Z_6X_7Z_8+Z_9X_{10}Z_{11}]+\epsilon X_{13}
\end{equation*}
and choose the vector $b$ as the 13-qubit cluster state 
\begin{equation*}
b= \ket{C_{13}}
=
\bigg(\prod_{i=1}^{12} CZ_{i,i+1}\bigg)
\bigg(\bigotimes_{i=1}^{13}\ket{+}_i\bigg)
=U_b\ket{0}^{\otimes 13},
\end{equation*}
where $c_1, c_2, \epsilon $ are real scalar coefficients, $X_{13}$ denotes the Pauli-$X$ operator acting on the 13th qubit and introduces a local perturbation, $CZ_{i,i+1}$ denotes the controlled-$Z$ gate acting on qubits $i$ and $i+1$, $\otimes$ denotes the tensor product, 
$\ket{+} = (\ket{0} + \ket{1})/\sqrt{2}$, $\ket{+}_i$ denotes the state $\ket{+}$ on the $i$th qubit, and
$U_b = ( \prod_{i=1}^{12} CZ_{i,i+1} ) H^{\otimes 13}$.
The multi-qubit terms correspond to cluster-state stabilizers of the form $Z_{i-1} X_i Z_{i+1}$, with boundary terms adapted accordingly.
This linear system is inspired by stabilizer codes in quantum error correction.
Since $\ket{b}$ is a common $+1$ eigenstate of the stabilizers $X_1Z_2$, $Z_3X_4Z_5$, $Z_6X_7Z_8$, and $Z_9X_{10}Z_{11}$, it is an eigenstate of the unperturbed matrix $A'=c_1 I +c_2[X_1Z_2+Z_3X_4Z_5+Z_6X_7Z_8+Z_9X_{10}Z_{11}]$. Therefore, for the unperturbed system $A'\ket{x}=\ket{b}$, the exact solution is $\ket{x}=(c_1+4c_2)^{-1}\ket{b}$.

In the simulations, the perturbation parameter $\epsilon$ is set to $0.1$, and the coefficients $c_1, c_2$ are tuned such that the condition number of $A$ is $20$.
The variational ansatz is implemented using a hardware-efficient circuit consisting of alternating layers of single-qubit $R_y$ rotations on each qubit and controlled-$Z$ gates acting on each nearest-neighbor qubit pair, with five layers of each type. The circuit parameters $\alpha_{ij}$ and $\beta_{ij}$ are initialized 
uniformly at random over $[-\pi,\pi]$, while the norm variables $\rho_{ij}$ and $\sigma_{ij}$ are initialized to $1$.
The cost functions $C_{ij}$ and their gradients are evaluated via simulations of the Hadamard-test circuits. The hybrid quantum-classical optimization is performed with base stepsize $\eta=0.01$. All reported results are averaged over 10 independent trials corresponding to different random initializations.


We measure convergence using the norm of the global residual $\|Ax(t) - b\|$, where the global estimate $x(t)$ is defined as follows. 
For an $m^2$-agent network corresponding to an $m \times m$ block partition, 
$x(t) = \frac{1}{m} \sum_{i=1}^m x_i(t)$, where 
$x_i=
[
   (\rho_{i1}\ket{x_{i1}})^\top,  
      \cdots, 
   (\rho_{im}\ket{x_{im}})^\top 
]^\top$.
This averaging across block rows and stacking across block columns ensures that the global estimate aggregates the information of all agents and converges to a global solution as consensus is achieved among block rows.
Accordingly, we define the consensus error as the standard deviation of $\{x_i(t)\}_{i=1}^m$ across different block rows, i.e.,
$
\sqrt{\frac{1}{m} \sum_{i=1}^m \|x_i(t) - x(t)\|^2}$.

\begin{figure}[!t]
    \centering
    \includegraphics[width=0.485\textwidth]{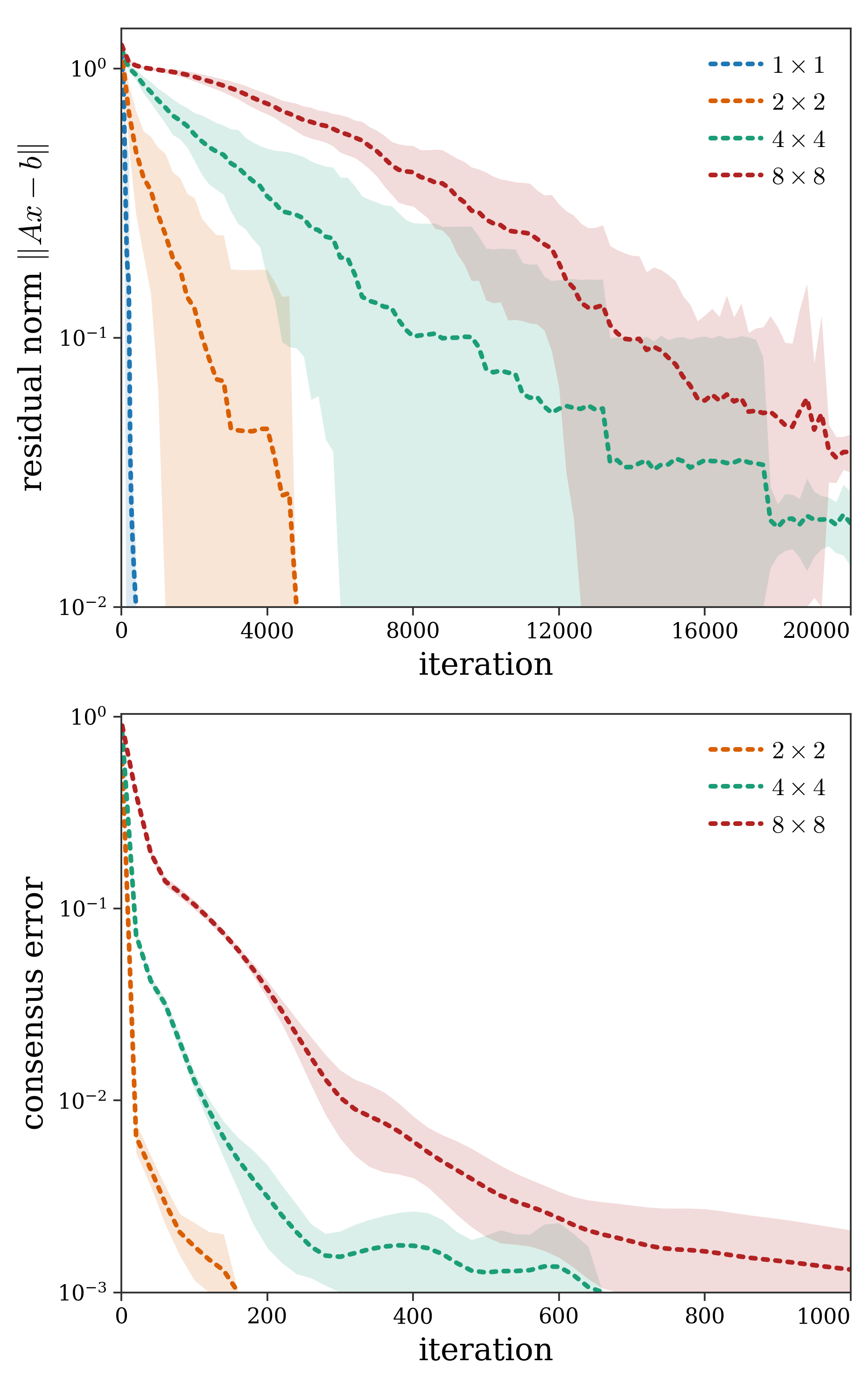}
    \caption{Global residual norm and consensus error performance comparison of the distributed VQLS algorithm under different block partitions for the $13$-qubit perturbed cluster-state stabilizer system}
    \label{fig:partition_comparsion}
\end{figure}

With the above settings, we compare four prefix block partitions of the same $13$-qubit linear system, namely $1\times1$, $2\times2$, $4\times4$, and $8\times8$. 
The $1\times1$ case corresponds to the single-agent setting that solves the linear system in a centralized manner,\footnote{
The proposed distributed VQLS algorithm reduces to a single-agent centralized VQLS-type algorithm when $m=1$; however, it differs from the VQLS algorithm in \cite{vqls}, as the cost function is defined differently to enable distributed implementation.
} 
while the latter three correspond to $4$-, $16$-, and $64$-agent networks, respectively, where the proposed distributed algorithm is employed. 
For each partition case, both the row and column neighbor graphs are chosen as minimally connected path graphs.
The trajectories of the global residual norm and the consensus error of the proposed distributed VQLS algorithm under these block partitions are shown in Figure \ref{fig:partition_comparsion}.
The figure shows that all partition cases converge to a common least squares solution. For the $4\times4$ and $8\times8$ block partitions, the global residual norm will continue to decrease as the number of iterations increases, but eventually fluctuates around $10^{-2}$. Further reduction in the residual can be achieved by tuning the base stepsize and/or adjusting the connectivity of the neighbor graphs.
Meanwhile, the consensus error reaches a similarly small level within significantly fewer iterations for all distributed cases, indicating that agreement among block rows can be achieved faster than convergence of the optimization.


The comparison among these partition cases shows that the (distributed) optimization procedure generally becomes slower as the partition becomes finer. The centralized $1\times1$ case exhibits the fastest convergence, while the distributed cases remain effective but tend to converge more slowly as the number of blocks increases. In particular, the $8\times8$ partition still achieves the target accuracy, although its convergence is the slowest within the same iteration budget.
A plausible explanation is that, as the partition grid becomes denser, each agent has access to a smaller local portion of the global linear system, so recovering the global solution requires more coordination among agents. 
This behavior is analogous to classical distributed optimization, where distributed algorithms typically converge more slowly than their centralized counterparts, although they can achieve the same convergence order under suitable conditions depending on the network size and connectivity \cite{nedic2018network}. 
From the results shown in the figure, it appears that, as the number of partitions increases, the convergence behavior deviates from that of the centralized single-agent case and may not exhibit a comparable convergence order. It is possible that achieving comparable convergence rates to the centralized setting is more difficult in the quantum regime, as the underlying optimization involves highly nonlinear and nonconvex quantum operations.
It is worth emphasizing that the primary goal of the proposed distributed framework is to solve larger-scale problems that may exceed the capability of a single quantum device, rather than to outperform the convergence behavior of the centralized approach, although achieving comparable performance would be desirable.

\subsection{Failure of Distributed Gradient Descent}\label{subsec:gradient_descent_fail}

Recall that the Adam optimization technique is used in the proposed algorithm among row-neighboring agents to update the parameter vector $\tilde\beta_{ij}$, which parametrizes the optimization variable $z_{ij}$, 
while both Adam and gradient tracking techniques are employed among column-neighboring agents to update the parameter vector $\tilde\alpha_{ij}$, which parametrizes the optimization variable $x_{ij}$. 
Both techniques can be simplified to gradient descent, which is known to perform well in classical distributed optimization, albeit with slower convergence. 
This subsection demonstrates that if either Adam or gradient tracking is replaced by gradient descent, the resulting distributed algorithm may fail to converge. 
To this end, we consider three variants of the proposed algorithm, each obtained by replacing one Adam or gradient tracking step with gradient descent, as described as follows: 

\vspace{.05in}

\noindent
{\bf Variant 1:}
The update of $\tilde\alpha_{ij}$ uses gradient descent instead of Adam while retaining gradient tracking, whereas the update of $\tilde\beta_{ij}$ continues to employ Adam. Accordingly, a variant of the proposed distributed algorithm is obtained by replacing the update step \eqref{eq:alpha_update} with 
    \begin{equation*}
        \tilde{\alpha}_{ij}(t+1)
        = \textstyle \sum_{k
    \in\mathcal M_{ij}} w_{ik}\tilde{\alpha}_{kj}(t)
        - \eta {y}_{ij}(t).
    \end{equation*}
This results in a delayed gradient-tracking DIGing update.
Consequently, the corresponding Adam related auxiliary update steps \eqref{eq:alpha_Adam1} and \eqref{eq:alpha_Adam2} are no longer needed.
This variant is labeled \textbf{Track+AdamZ}  in the plot.

\vspace{.05in}

\noindent
{\bf Variant 2:}
The update of $\tilde\beta_{ij}$ uses gradient descent instead of Adam, while the update of $\tilde\alpha_{ij}$ continues to employ both Adam and gradient tracking.
Accordingly, a variant of the proposed distributed algorithm is obtained by replacing the Adam update step \eqref{eq:beta_update} with 
    \begin{equation*}
        \tilde{\beta}_{ij}(t+1)
        = \tilde{\beta}_{ij}(t) - \eta {g}_{ij}(t).
    \end{equation*}
Consequently, the corresponding Adam related auxiliary update steps \eqref{eq:beta_Adam1} and \eqref{eq:beta_Adam2} are no longer needed.
This variant is labeled \textbf{Track+AdamX} in the plot.

\vspace{.05in}

\noindent
{\bf Variant 3:}
The update of $\tilde\alpha_{ij}$ removes gradient tracking while retaining Adam,\footnote{
Replacing gradient tracking with gradient descent while retaining Adam results in an Adam-type update.
} whereas the update of $\tilde\beta_{ij}$ continues to employ Adam. 
Accordingly, a variant of the proposed distributed algorithm is obtained by 
removing the gradient tracking step \eqref{eq:tracker_y_update}
and replacing the steps \eqref{eq:alpha_Adam1} and \eqref{eq:alpha_Adam2} with 
    \begin{align*}
        \mu_{ij}(t+1)
        &= \gamma_1 \mu_{ij}(t)
        + (1-\gamma_1)g'_{ij}(t), \\
        \nu_{ij}(t+1)
        &= \gamma_2 \nu_{ij}(t)
        + (1-\gamma_2)(g'_{ij}(t))^{\odot 2},
    \end{align*}
where $g'_{ij}(t) = \nabla_{\tilde\alpha_{ij}}C_{ij}(t)$. 
This yields a consensus-based distributed Adam update for $\tilde\alpha_{ij}$.
This variant is labeled \textbf{Consensus+AdamX+AdamZ} in the plot.

\vspace{.05in}

To evaluate the performance of the three variant algorithms, we consider the Ising-inspired linear system given by~\eqref{eq:Ising_A}--\eqref{eq:Ising_b} with $n=7$, and adopt the $4\times4$ block partition, corresponding to a $16$-agent network, on which all three variants, along with the proposed algorithm, are implemented. 
The simulation setup is as follows. Both the row- and column-neighbor graphs are chosen to be the path graph. The coupling parameter is set to $\kappa=0.1$, and the other coefficients are selected such that the condition number of $A$ is $200$. Each agent employs the same variational ansatz, consisting of three alternating layers of single-qubit $R_y$ rotations on each qubit and controlled-$Z$ gates acting on each nearest-neighbor qubit pair. The parameters in $\alpha_{ij}$ and $\beta_{ij}$ are randomly initialized in $[-\pi,\pi]$, while $\rho_{ij}$ and $\sigma_{ij}$ are initialized to $1$. The cost function is evaluated via simulated Hadamard test circuits. All simulations use a common (base) stepsize $\eta=0.01$, and the reported results are averaged over $10$ random initializations.

\begin{figure}[!ht]
    \centering
    \includegraphics[width=0.485\textwidth]{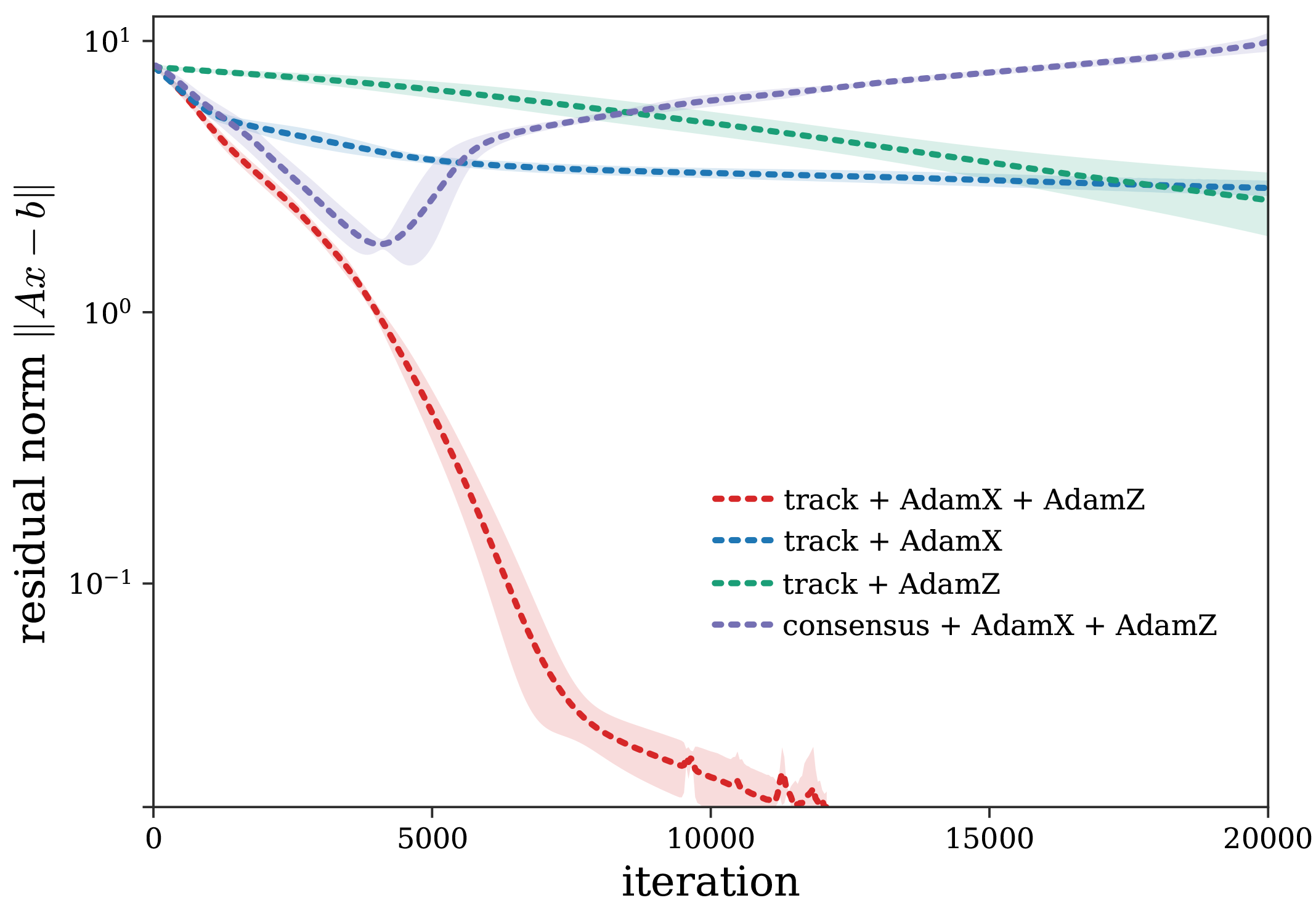}
    \caption{Global residual norm trajectories of the proposed distributed VQLS algorithm and its three variants applied to the 7-qubit Ising-inspired linear system in \eqref{eq:Ising_A} and \eqref{eq:Ising_b} with a $4 \times 4$ block partition and a 16-agent network}
\vspace{-.15in}    
    \label{fig:necessarity_comp}
\end{figure}

With the above setup, the global residual norm trajectories of the proposed distributed VQLS algorithm and its three variants are shown in Figure \ref{fig:necessarity_comp}, in which the proposed algorithm is labeled as \textbf{Track+AdamX+AdamZ}.
From the figure, among the four algorithms, only the proposed algorithm exhibits a sustained and rapid decrease in the global residual norm. The residual decreases by more than two orders of magnitude from its initial value within approximately $10^4$ iterations. In contrast, the other three variants exhibit significantly slower progress and remain above unity even after $2\times10^4$ iterations. 
These results indicate that, in this setting, only the proposed algorithm converges reliably, while all three variants fail to reach a satisfactory solution.
This observation highlights that both gradient tracking and Adam updates play a critical role in the practical performance of the proposed distributed VQLS algorithm.

\section{Conclusion}

This paper proposed a distributed variational quantum algorithm for solving large-scale linear systems, where the system matrix is partitioned across multiple NISQ devices with limited local information. By integrating a tailored variational quantum linear solver at each agent with distributed classical optimization techniques over connected row- and column-neighbor graphs, the framework enables cooperative computation of a global least squares solution through local quantum evaluations and classical information exchange. The design of the quantum cost function plays a central role in establishing the distributed structure and guiding the algorithm development. Numerical simulations validate that the proposed approach achieves scalability with respect to the number of participating quantum devices, thereby overcoming the size limitations inherent to individual NISQ processors. These results highlight the potential of distributed hybrid quantum-classical architectures for solving large-scale computational problems beyond the capability of single-device quantum systems.

While the proposed framework demonstrates the feasibility of distributed variational quantum algorithms for large-scale linear systems, it represents only a promising initial step, and several important directions remain to be explored. On the numerical side, more extensive simulations are needed to better understand the scalability of the algorithm as the number of agents increases, as well as its performance under varying condition numbers, similar to the analyses conducted in~\cite{vqls}. It is also of interest to systematically investigate how the connectivity properties of the row- and column-neighbor graphs affect convergence behavior. On the algorithmic side, an important direction is to further reduce the communication overhead, in particular by designing schemes that require only a single transmission round per iteration. In addition, extending the framework to account for communication delays and asynchronous updates is essential for practical deployment in distributed quantum-classical architectures.

The most important next step is to implement and evaluate the proposed algorithm on real quantum hardware, such as IBM Quantum devices. The current simulation environment does not fully capture realistic noise sources, including device-dependent gate errors, decoherence, and readout noise. By distributing the computation across multiple agents, the proposed framework reduces the quantum circuit depth executed on each device, which may improve robustness to noise. It is therefore of significant interest to experimentally assess whether, as expected, the distributed algorithm exhibits enhanced noise resilience compared to single device implementations. Such validation is essential for understanding the practical advantages and limitations of distributed variational quantum algorithms in the NISQ era.


\section*{Appendix}

\subsection{Proof of Lemma \ref{lemma:goal_fixed}}\label{append:proof}

For any \(x\) and  \(i\in\{1,\ldots,m\}\), define
$
r_i(x)=\bar A_i x-\bar b_i$.
Since \(\bbb G^{{\rm row}}\) is connected, its Laplacian matrix \(L\) satisfies
$
\Ker(L)=\mathrm{span}\{\1_m\}$,
where $\1_m$ denotes the $m$-dimensional vector with all entries equal to $1$.
Since \(L\) is symmetric, 
$
\Ker(\bar L)=\Ker(\bar L^\top)
=\{\1_m\otimes v: v\in\R^{2^q}\}$.
Now fix \(x\) and \(i\).
Minimizing over \(z_i\) yields the distance from \(r_i(x)\) to the image of \(\bar L\), hence
\begin{equation}\label{eq:proj_identity_vqls}
\min_{z_i}\|r_i(x)-\bar L z_i\|^2
=
\|P_{\Ker(\bar L^\top)}r_i(x)\|^2,
\end{equation}
where $P_{\Ker(\bar L^\top)}$ denotes the orthogonal projector onto the subspace $\Ker(\bar L^\top)$. Note that
$
P_{\Ker(\bar L^\top)}(y)
=
(\1_m\otimes I)(\frac1m(\1_m^\top\otimes I)y)$ for all $y\in\R^{m2^q}$.
Applying this to \(y=r_i(x)\), 
\[
P_{\Ker(\bar L^\top)}r_i(x)
=
(\1_m\otimes I)\textstyle(\frac1m(\1_m^\top\otimes I)(\bar A_i x-\bar b_i)).
\]
Since 
$
(\1_m^\top\otimes I)\bar A_i x=A_i x$ and 
$(\1_m^\top\otimes I)\bar b_i=b_i$,
$
(\1_m^\top\otimes I)(\bar A_i x-\bar b_i)=A_i x-b_i$.
Then,
\[
P_{\Ker(\bar L^\top)}r_i(x)
=
(\1_m\otimes I)\textstyle(\frac1m(A_i x-b_i)).
\]
Substituting this into \eqref{eq:proj_identity_vqls} yields
\[
\min_{z_i}\|\bar A_i x-\bar b_i-\bar L z_i\|^2
=
\textstyle\|(\1_m\otimes I)\frac1m(A_i x-b_i)\|^2.
\]
Using \(\|\1_m\otimes v\|^2=\|\1_m\|^2\|v\|^2=m\|v\|^2\), we conclude that
$
\min_{z_i}\|\bar A_i x-\bar b_i-\bar L z_i\|^2
=
\frac1m\|A_i x-b_i\|^2$.
Thus,
\begin{align}
& \textstyle\min_{\{z_i\}_{i=1}^m}\textstyle\sum_{i=1}^m\|\bar A_i x-\bar b_i-\bar L z_i\|^2 \nonumber\\
=\;& \textstyle
\frac1m\sum_{i=1}^m\|A_i x-b_i\|^2
=
\frac1m\|Ax-b\|^2. \label{eq:reduced_objective_vqls}
\end{align}

We first prove the sufficiency. 
Suppose that \((x^*,\{z_i^*\})\) minimizes \(F(x,\{z_i\})=\sum_{i=1}^m\|\bar A_i x-\bar b_i-\bar L z_i\|^2\). Define \(g(x)=\min_{\{z_i\}}F(x,\{z_i\})\). Then, \(g(x^*)\le F(x^*,\{z_i^*\})\le F(x,\{z_i\})\) for all \(x\) and \(\{z_i\}\). Minimizing the right hand side over \(\{z_i\}\) yields \(g(x^*)\le g(x)\) for all \(x\). It follows that \(x^*\) minimizes the left hand side of \eqref{eq:reduced_objective_vqls}, and thus minimizes \(\|Ax-b\|^2\). Therefore, \(x^*\) is a least squares solution to \(Ax=b\).
Conversely, to establish necessity, suppose that \(x^*\) is a least squares solution to \(Ax=b\).
For each \(i\), choose \(z_i^*\) to be any minimizer of
$
\min_{z_i}\|\bar A_i x^*-\bar b_i-\bar L z_i\|^2$.
Such a minimizer exists because \(\operatorname{Im}(\bar L)\) is a linear subspace of \(\R^{m2^q}\), so the minimum distance from \(\bar A_i x^*-\bar b_i\) to \(\operatorname{Im}(\bar L)\) is attained by orthogonal projection. Then, from \eqref{eq:reduced_objective_vqls},
$
F(x^*,\{z_i^*\})
=
\frac1m\|Ax^*-b\|^2$.
Since \(x^*\) minimizes \(\|Ax-b\|^2\), \eqref{eq:reduced_objective_vqls} implies that \(F(x^*,\{z_i^*\})\le F(x,\{z_i\})\) for all \(x\) and \(\{z_i\}\). Hence, \((x^*,\{z_i^*\})\) minimizes \(F(x,\{z_i\})\).

\subsection{Hadamard Test Circuits}

\begin{figure}[!t]
    \centering
    \includegraphics[width=0.485\textwidth]{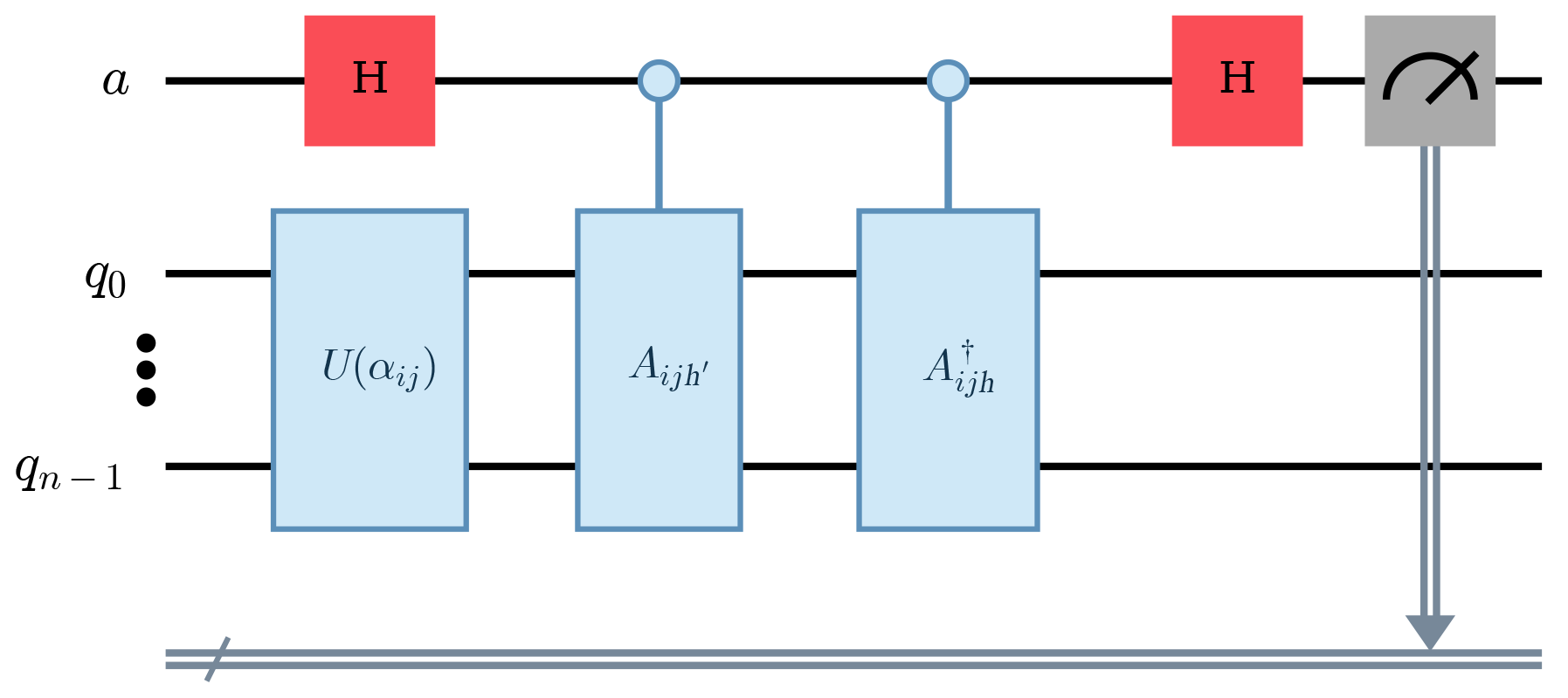}
    \caption{Hadamard test circuit for \eqref{eq:overlap_test}
    }
    \label{fig:overlap_test}
\end{figure}

\begin{figure}[!t]
    \centering
    \includegraphics[width=0.485\textwidth]{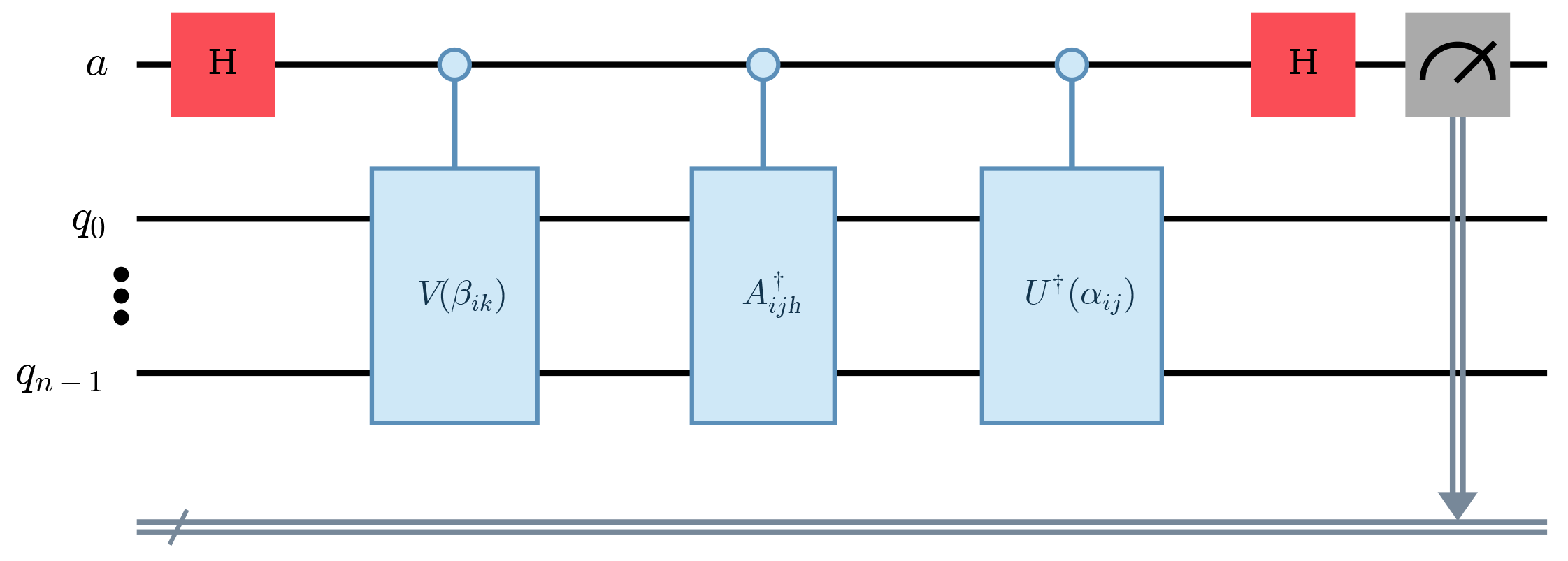}
    \caption{Hadamard test circuit for \eqref{eq:hadamard1}
    }
    \label{fig:hadamard1}
\end{figure}

\begin{figure}[!t]
    \centering
    \includegraphics[width=0.485\textwidth]{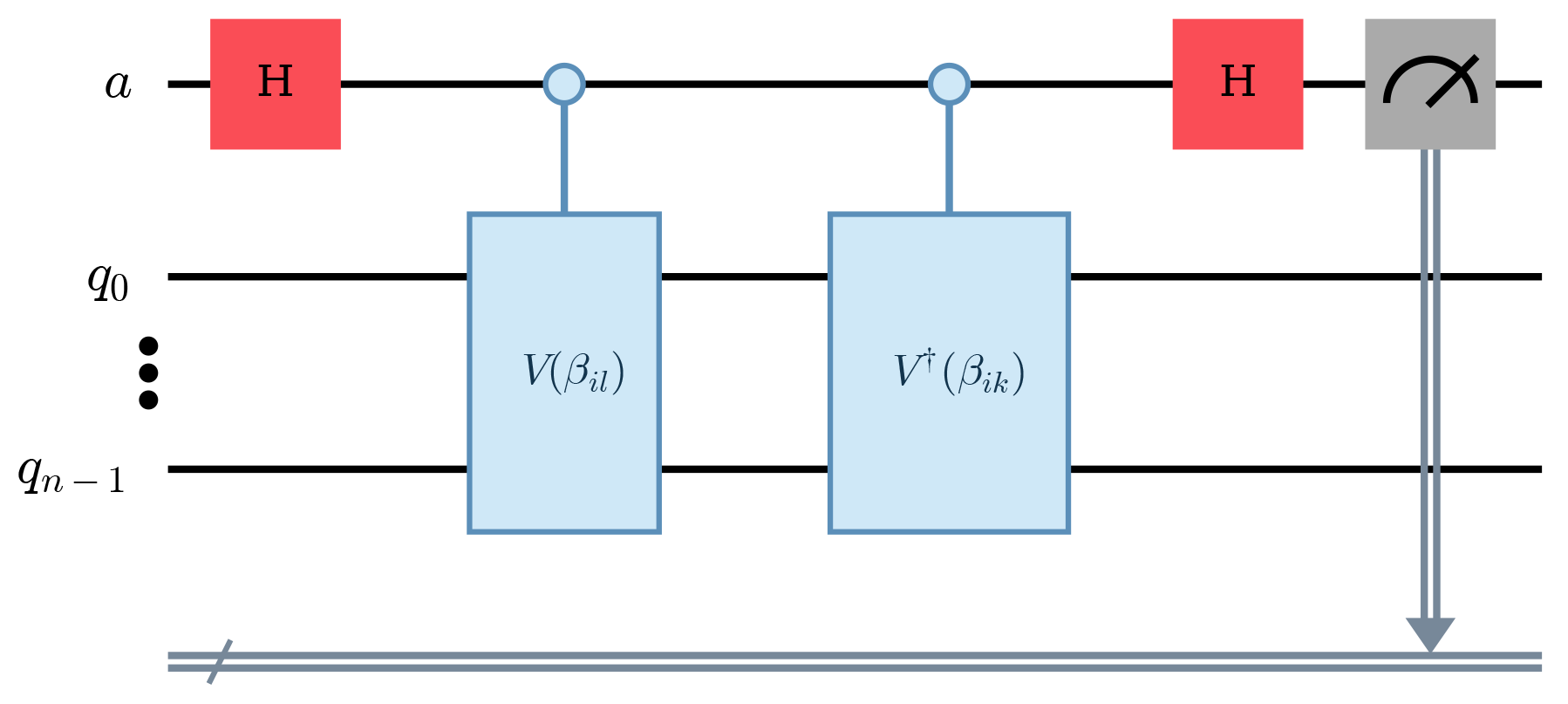}
    \caption{Hadamard test circuit for \eqref{eq:hadamard2}
    }
    \label{fig:hadamard2}
\end{figure}

\begin{figure}[!t]
    \centering
    \includegraphics[width=0.485\textwidth]{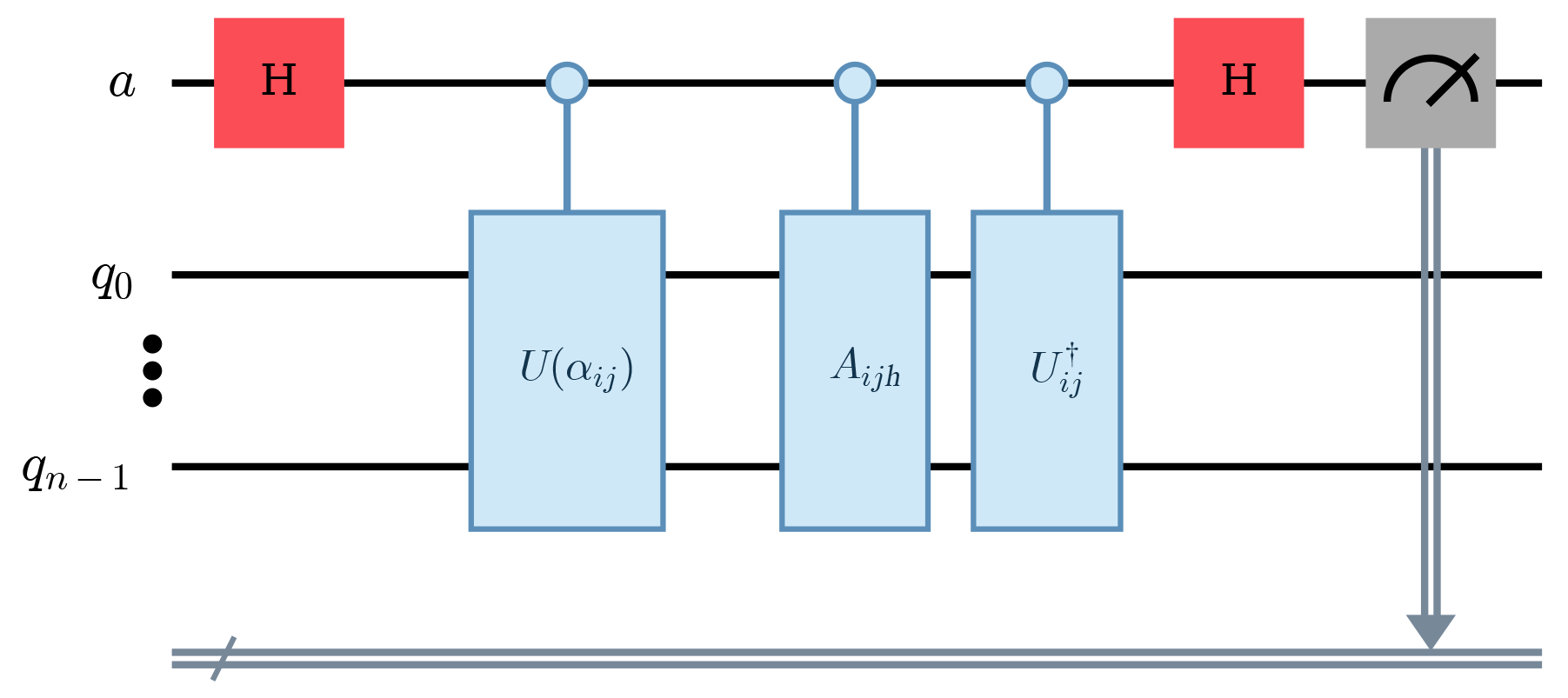}
    \caption{Hadamard test circuit for \eqref{eq:hadamard3}
    }
    \label{fig:hadamard3}
\end{figure}

\begin{figure}[!t]
    \centering
    \includegraphics[width=0.485\textwidth]{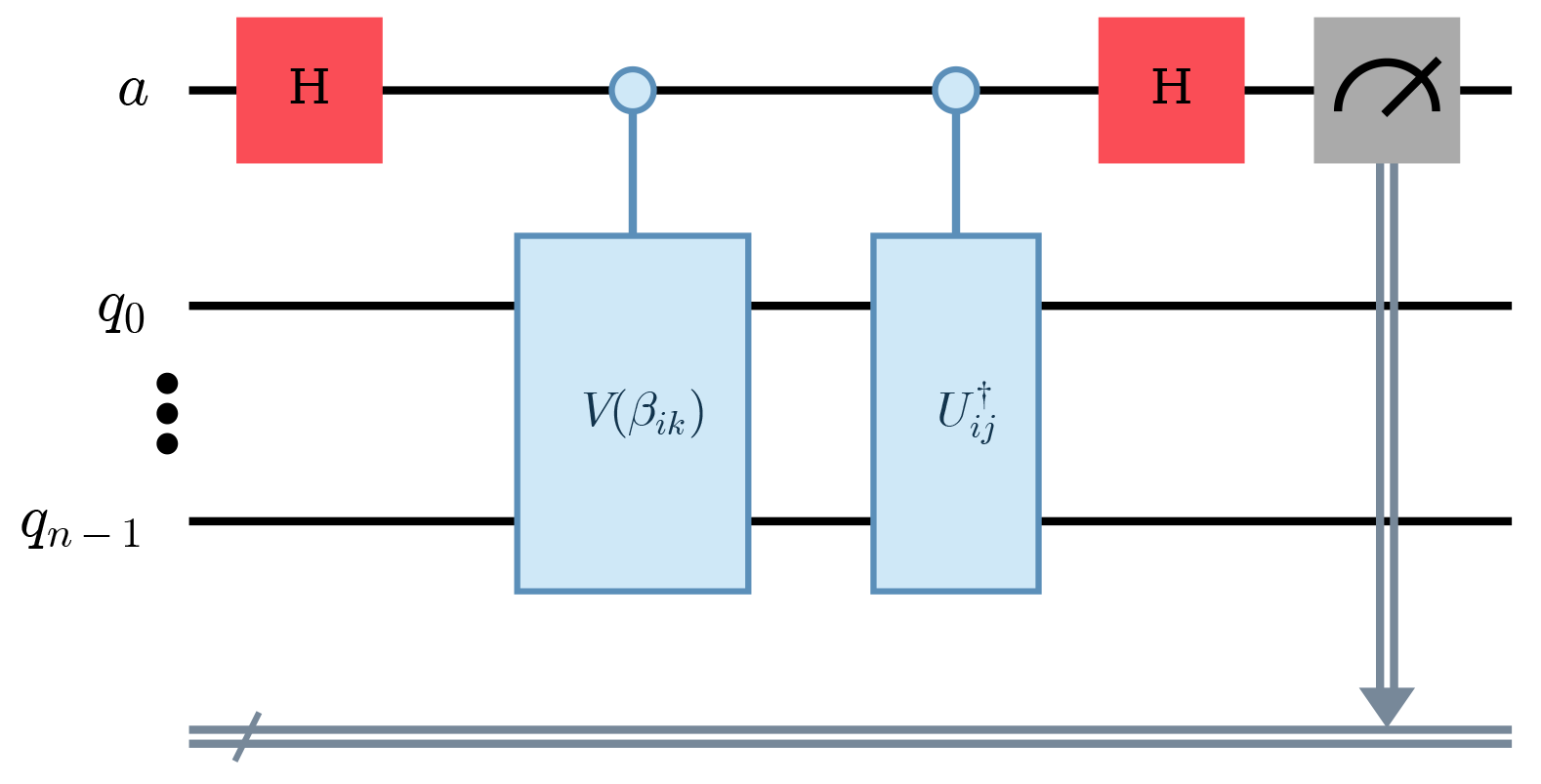}
    \caption{Hadamard test circuit for \eqref{eq:hadamard4}
    }
    \label{fig:hadamard4}
\end{figure}

This subsection presents 
the Hadamard test circuits for the inner product terms in \eqref{eq:overlap_test}--\eqref{eq:hadamard4}, as shown in Figures \ref{fig:overlap_test}--\ref{fig:hadamard4}.
In these circuit diagrams, $a$ denotes the ancilla qubit, and $q_0, \ldots, q_{n-1}$ denote the $n$ system qubits at each agent; the operators $A_{ijh}$ and $A_{ijh'}$ are the $h$th and $h'$th unitary terms in the LCU decomposition of $A_{ij}$.

\begin{figure*}
\hrule
\begin{align}
    \textstyle \pdv{\alpha_{ijh}} \mel{x_{ij}}{A_{ij}^\dagger A_{ij}}{x_{ij}} &= \textstyle\frac{1}{2} \big[ \mel{0}{U_+^\dagger(\alpha_{ijh}) A_{ij}^\dagger A_{ij} U_+(\alpha_{ijh})}{0} - \mel{0}{U_-^\dagger(\alpha_{ijh}) A_{ij}^\dagger A_{ij} U_-(\alpha_{ijh})}{0} \big], \label{eq:gradient1}\\
    \textstyle \pdv{\alpha_{ijh}} \mel{x_{ij}}{A_{ij}^\dagger}{z_{ik}} &= \textstyle \frac{1}{2\sqrt{2}} \big[ \mel{0}{U_+^\dagger(\alpha_{ijh}) A_{ij}^\dagger V(\beta_{ik})}{0} - \mel{0}{U_-^\dagger(\alpha_{ijh}) A_{ij}^\dagger V(\beta_{ik})}{0} \big], \\
    \textstyle \pdv{\beta_{ikh}} \mel{x_{ij}}{A_{ij}^\dagger}{z_{ik}} &= \textstyle \frac{1}{2\sqrt{2}} \big[ \mel{0}{U^\dagger(\alpha_{ij}) A_{ij}^\dagger V_+(\beta_{ikh})}{0} - \mel{0}{U^\dagger(\alpha_{ij}) A_{ij}^\dagger V_-(\beta_{ikh})}{0} \big], \\
    \textstyle \pdv{\beta_{ikh}} \braket{z_{ik}}{z_{il}} &= \textstyle \frac{1}{2\sqrt{2}} \big[ \mel{0}{V_+^\dagger(\beta_{ikh}) V(\beta_{il})}{0} - \mel{0}{V_-^\dagger(\beta_{ikh}) V(\beta_{il})}{0} \big], \\
    \textstyle \pdv{\beta_{ilh}} \braket{z_{ik}}{z_{il}} &= \textstyle \frac{1}{2\sqrt{2}} \big[ \mel{0}{V^\dagger(\beta_{ik}) V_+(\beta_{ilh})}{0} - \mel{0}{V^\dagger(\beta_{ik}) V_-(\beta_{ilh})}{0} \big], \\
    \textstyle \pdv{\alpha_{ijh}} \mel{b_{ij}}{A_{ij}}{x_{ij}} &= \textstyle \frac{1}{2\sqrt{2}} \big[ \mel{0}{U_{ij}^\dagger A_{ij} U_+(\alpha_{ijh})}{0} - \mel{0}{U_{ij}^\dagger A_{ij} U_-(\alpha_{ijh})}{0} \big], \\
    \textstyle \pdv{\beta_{ikh}} \braket{b_{ij}}{z_{ik}} &= \textstyle \frac{1}{2\sqrt{2}} \big[ \mel{0}{U_{ij}^\dagger V_+(\beta_{ikh})}{0} - \mel{0}{U_{ij}^\dagger V_-(\beta_{ikh})}{0} \big]. 
    \label{eq:gradient7}
\end{align}
\hrule
\end{figure*}

\subsection{Cost Function Gradients}

Based on the expression of the cost function $C_{ij}$ and its detailed decomposition in the Cost Function part in Subsection \ref{subsec:algorithm}, the gradients of $C_{ij}$ with respect to the parameter vectors $\alpha_{ij}$ and $\beta_{ij}$ reduce to computing the gradients of the constituent terms in \eqref{eq:overlap_test}--\eqref{eq:hadamard4}. The resulting gradient expressions, derived via the parameter shift rule, are given in \eqref{eq:gradient1}--\eqref{eq:gradient7}, where 
\begin{align*}
    U_{+}(\alpha_{ijh}) &= G_{p}(\alpha_{ijp}) \cdots G_{h}(\alpha_{ijh} + \textstyle\frac{\pi}{2}) \cdots G_{1}(\alpha_{ij1}), \\
    U_{-}(\alpha_{ijh}) &= G_{p}(\alpha_{ijp}) \cdots G_{h}(\alpha_{ijh} - \textstyle\frac{\pi}{2}) \cdots G_{1}(\alpha_{ij1}), \\
    V_{+}(\alpha_{ijh}) &= H_{q}(\alpha_{ijq}) \cdots H_{h}(\alpha_{ijh} + \textstyle\frac{\pi}{2}) \cdots H_{1}(\alpha_{ij1}), \\
    V_{-}(\alpha_{ijh}) &= H_{q}(\alpha_{ijq}) \cdots H_{h}(\alpha_{ijh} - \textstyle\frac{\pi}{2}) \cdots H_{1}(\alpha_{ij1}).
\end{align*}
Note that terms depending on multiple parameter vectors (e.g., $\alpha_{ij}$ and $\beta_{ik}$, or $\beta_{ik}$ and $\beta_{i\ell}$) yield multiple corresponding gradient expressions.

Since the expressions in \eqref{eq:gradient1}--\eqref{eq:gradient7} contain inner product terms similar to those in \eqref{eq:overlap_test}--\eqref{eq:hadamard4}, they can be evaluated using the same types of Hadamard test 
circuits designed for \eqref{eq:overlap_test}--\eqref{eq:hadamard4}.
The cost function gradients also involve derivatives with respect to the real-valued parameters $\rho_{ij}$ and $\sigma_{ij}$, which are easy to compute without quantum evaluations and hence omitted here.


\section*{Acknowledgement}

The authors are grateful to Hrushikesh Pramod Patil, Eddie Chen, and Yiming Zeng for their efforts during preliminary investigations of this problem a few years ago, and wish to thank Tzu-Chieh Wei and Nhat Anh Nghiem for their useful discussions.

\bibliographystyle{unsrt}
\bibliography{jicareer,quantum}

@article{tacle,
    title     = "A Distributed Algorithm for Solving a Linear Algebraic Equation",
    author    = "S. Mou and J. Liu and A.S. Morse",
    journal   = "IEEE Transactions on Automatic Control",
    year      = "2015",
    volume    = "60",
    number    = "11",
    pages     = "2863-2878"
}

@article{vqls,
  title   = {Variational Quantum Linear Solver},
  author  = {C. Bravo-Prieto and R. LaRose and M. Cerezo and Y. Suba{\c{s}}{\i} and L. Cincio and P.J. Coles},
  journal = {Quantum},
  volume  = {7},
  pages   = {1188},
  year    = {2023},
}

@inproceedings{grover1996fast,
  title={A fast quantum mechanical algorithm for database search},
  author={L.K. Grover},
  booktitle={Proceedings of the 28th Annual ACM Symposium on Theory of Computing},
  pages={212--219},
  year={1996}
}

@article{mcclean2016theory,
  title   = {The theory of variational hybrid quantum-classical algorithms},
  author  = {J.R. McClean and J. Romero and R. Babbush and A. Aspuru-Guzik},
  journal = {New Journal of Physics},
  volume  = {18},
  pages   = {023023},
  year    = {2016},
}

@article{molzahn2017survey,
  title={A survey of distributed optimization and control algorithms for electric power systems},
  author={Molzahn, D.K. and D{\"o}rfler, F. and Sandberg, H. and Low, S.H. and Chakrabarti, S. and Baldick, R. and Lavaei, J.},
  journal={IEEE Transactions on Smart Grid},
  volume={8},
  number={6},
  pages={2941--2962},
  year={2017},
}

@article{circuit_partitioning,
  title = {Automated distribution of quantum circuits via hypergraph partitioning},
  author = {P. Andr\'es-Mart\'{\i}nez and C. Heunen},
  journal = {Physical Review A},
  volume = {100},
  issue = {3},
  pages = {032308},
  numpages = {11},
  year = {2019},
  publisher = {American Physical Society},
  doi = {10.1103/PhysRevA.100.032308},
  url = {https://link.aps.org/doi/10.1103/PhysRevA.100.032308}
}

@article{wang2020scalable,
  author  = {X. Wang and S. Mou and B.D.O. Anderson},
  title   = {Scalable, Distributed Algorithms for Solving Linear Equations via Double-Layered Networks},
  journal = {IEEE Transactions on Automatic Control},
  volume  = {65},
  number  = {3},
  pages   = {1132--1143},
  year    = {2020},
}

@article{huang2022scalable,
  author  = {Y. Huang and Z. Meng and J. Sun},
  title   = {Scalable distributed least squares algorithms for large-scale linear equations via an optimization approach},
  journal = {Automatica},
  volume  = {146},
  pages   = {110572},
  year    = {2022},
}

@article{huang2024distributed,
  title={Distributed algorithms for solving a least-squares solution of linear algebraic equations},
  author={Y. Huang and Z. Meng},
  journal={IEEE Transactions on Control of Network Systems},
  volume={11},
  number={2},
  pages={599--609},
  year={2024}
}

@inproceedings{pham2023distributed,
  author    = {V.H. Pham and H.-S. Ahn},
  title     = {Distributed Least Square Approach for Solving a Multiagent Linear Algebraic Equation},
  booktitle = {Proceedings of the 62nd IEEE Conference on Decision and Control},
  year      = {2023},
  pages     = {7259--7264}
}

@inproceedings{cao2017continuous,
  author    = {K. Cao and X. Zeng and Y. Hong},
  title     = {Continuous-Time Distributed Algorithms for Solving Linear Algebraic Equations},
  booktitle = {Proceedings of the 36th Chinese Control Conference},
  year      = {2017},
  pages     = {8068--8073}
}

@article{schuld2019evaluating,
  title={Evaluating analytic gradients on quantum hardware},
  author={M. Schuld and V. Bergholm and C. Gogolin and J. Izaac and N. Killoran},
  journal={Physical Review A},
  volume={99},
  number={3},
  pages={032331},
  year={2019}
}

@article{gtadam,
  title={{GTA}dam: gradient Tracking With Adaptive Momentum for Distributed Online Optimization},
  author={G. Carnevale and F. Farina and I. Notarnicola and G. Notarstefano},
  journal={IEEE Transactions on Control of Network Systems},
  volume={10},
  number={3},
  pages={1436--1448},
  year={2023}
}

@inproceedings{metro2,
    author    = "L. Xiao and S. Boyd and S. Lall",
    title     = "A scheme for robust distributed sensor fusion based on average consensus",
    year      = "2005",
    booktitle = "Proceedings of the 4th International Conference on Information Processing in Sensor Networks",
    pages     = "63--70"
}

@article{du2022tqe,
  title={A Distributed Learning Scheme for Variational Quantum Algorithms},
  author={Y. Du and Y. Qian and X. Wu and D. Tao},
  journal={IEEE Transactions on Quantum Engineering},
  volume={3},
  pages={1--16},
  year={2022}
}

@article{PFEUTY197079,
title = {The one-dimensional {I}sing model with a transverse field},
journal = {Annals of Physics},
volume = {57},
number = {1},
pages = {79--90},
year = {1970},
issn = {0003-4916},
doi = {https://doi.org/10.1016/0003-4916(70)90270-8},
url = {https://www.sciencedirect.com/science/article/pii/0003491670902708},
author = {P. Pfeuty},
}

@phdthesis{gottesman,
  author       = {D. Gottesman},
  title        = {Stabilizer codes and quantum error correction},
  school       = {California Institute of Technology},
  year         = {1997},
}

@article{nedic2018network,
  title={Network topology and communication-computation tradeoffs in decentralized optimization},
  author={Nedi{\'c}, A. and Olshevsky, A. and Rabbat, M.G.},
  journal={Proceedings of the IEEE},
  volume={106},
  number={5},
  pages={953--976},
  year={2018},
  publisher={IEEE}
}

@INPROCEEDINGS{01_shor1994factorization,
  author={P.W. Shor},
  booktitle={Proceedings of the 35th Annual Symposium on Foundations of Computer Science}, 
  title={Algorithms for quantum computation: discrete logarithms and factoring}, 
  year={1994},
  pages={124--134},
}

@article{02_arute2019quantum,
  title={Quantum supremacy using a programmable superconducting processor},
  author={F. Arute and K. Arya and R. Babbush and others},
  journal={Nature},
  volume={574},
  pages={505--510},
  year={2019},
}

@misc{03_IBMQuantumRoadmap2023,
  author       = {{IBM Quantum}},
  title        = {The {IBM} Quantum Development Roadmap},
  year         = {2022},
  howpublished = {\url{https://www.ibm.com/quantum/roadmap}},
}

@article{04_liu2024creation,
  title={Creation of memory-memory entanglement in a metropolitan quantum network},
  author={J.L. Liu and X.Y. Luo and Y. Yu and others},
  journal={Nature},
  volume={629},
  pages={579--585},
  year={2024},
}

@article{05_yu2020entanglement,
  title={Entanglement of two quantum memories via fibres over dozens of kilometres},
  author={Y. Yu and F. Ma and X.Y. Luo and others},
  journal={Nature},
  volume={578},
  pages={240--245},
  year={2020},
}

@article{3_harrow2009hhl,
  title   = {Quantum Algorithm for Linear Systems of Equations},
  author  = {A.W. Harrow and A. Hassidim and S. Lloyd},
  journal = {Physical Review Letters},
  volume  = {103},
  number  = {15},
  pages   = {150502},
  numpages = {4},
  year    = {2009},
}

@article{5_childs2017precision,
  title   = {Quantum Algorithm for Systems of Linear Equations with Exponentially Improved Dependence on Precision},
  author  = {A.M. Childs and R. Kothari and R.D. Somma},
  journal = {SIAM Journal on Computing},
  volume  = {46},
  number  = {6},
  pages   = {1920--1950},
  year    = {2017},
}

@inproceedings{6_gilyen2019qsvt,
  title     = {Quantum Singular Value Transformation and Beyond: exponential Improvements for Quantum Matrix Arithmetics},
  author    = {A. Gily\'en and Y. Su and G.H. Low and N. Wiebe},
  booktitle = {Proceedings of the 51st Annual ACM SIGACT Symposium on Theory of Computing},
  pages     = {193--204},
  year      = {2019},
}

@article{10_preskill2018nisq,
  title   = {Quantum Computing in the {NISQ} Era and Beyond},
  author  = {J. Preskill},
  journal = {Quantum},
  volume  = {2},
  pages   = {79},
  year    = {2018},
}

@article{22_barral2025review,
  title   = {Review of Distributed Quantum Computing: from Single {QPU} to High Performance Quantum Computing},
  author  = {D. Barral and F.J. Cardama and G. D\'iaz-Camacho and others},
  journal = {Computer Science Review},
  volume  = {57},
  pages   = {100747},
  year    = {2025},
  doi     = {10.1016/j.cosrev.2025.100747},
  url     = {https://doi.org/10.1016/j.cosrev.2025.100747}
}

@article{24_chen2021fqml,
  title   = {Federated Quantum Machine Learning},
  author  = {S.Y.-C. Chen and S. Yoo},
  journal = {Entropy},
  volume  = {23},
  number  = {4},
  pages   = {460},
  year    = {2021},
  doi     = {10.3390/e23040460},
  url     = {https://doi.org/10.3390/e23040460}
}

@inproceedings{25_chehimi2021qfl,
  title     = {Federated Learning with Quantum Data},
  author    = {M. Chehimi and W. Saad},
  booktitle = {Proceedings of the 2022 IEEE International Conference on Acoustics, Speech and Signal Processing},
  pages     = {8617--8621},
  year      = {2022},
  doi       = {10.1109/ICASSP43922.2022.9746622},
  url       = {https://doi.org/10.1109/ICASSP43922.2022.9746622}
}

@inproceedings{26_chen2024cdqkl,
  title     = {Consensus-Based Distributed Quantum Kernel Learning for Speech Recognition},
  author    = {Chen, K.-C. and Ma, W. and Xu, X.},
  booktitle = {Proceedings of the 2025 IEEE International Conference on Acoustics, Speech, and Signal Processing Workshops},
  pages     = {1--5},
  year      = {2025},
  doi       = {10.1109/ICASSPW65056.2025.11011218},
  url       = {https://doi.org/10.1109/ICASSPW65056.2025.11011218}
}

@article{27_swaminathan2024securekernel,
title={Distributed and Secure Kernel-Based Quantum Machine Learning},
author={A. Swaminathan and M. Akg{\"u}n},
journal={Transactions on Machine Learning Research},
issn={2835-8856},
year={2025},
url={https://openreview.net/forum?id=3jdI0aEW3k},
}

@article{29_peng2020simulating,
  title   = {Simulating Large Quantum Circuits on a Small Quantum Computer},
  author  = {T. Peng and A.W. Harrow and M. Ozols and X. Wu},
  journal = {Physical Review Letters},
  volume  = {125},
  number  = {15},
  pages   = {150504},
  year    = {2020},
  doi     = {10.1103/PhysRevLett.125.150504},
  url     = {https://doi.org/10.1103/PhysRevLett.125.150504}
}

@article{34_nedic2017diging,
  title   = {Achieving Geometric Convergence for Distributed Optimization over Time-Varying Graphs},
  author  = {A. Nedi\'c and A. Olshevsky and W. Shi},
  journal = {SIAM Journal on Optimization},
  volume  = {27},
  number  = {4},
  pages   = {2597--2633},
  year    = {2017},
  doi     = {10.1137/16M1084316},
  url     = {https://doi.org/10.1137/16M1084316}
}

@inproceedings{41_kingma2015adam,
  title     = {Adam: a Method for Stochastic Optimization},
  author    = {D.P. Kingma and J. Ba},
  booktitle = {Proceedings of the 3rd International Conference on Learning Representations},
  year      = {2015},
  url       = {https://arxiv.org/abs/1412.6980}
}

@article{43_Yigit2019adiabatic,
  title = {Quantum Algorithms for Systems of Linear Equations Inspired by Adiabatic Quantum Computing},
  author = {Y. Suba\c{s}\i and R.D. Somma and D. Orsucci},
  journal = {Physical Review Letters},
  volume = {122},
  issue = {6},
  pages = {060504},
  numpages = {5},
  year = {2019},
  publisher = {American Physical Society},
  doi = {10.1103/PhysRevLett.122.060504},
  url = {https://link.aps.org/doi/10.1103/PhysRevLett.122.060504}
}

\end{document}